\pdfoutput=1

\documentclass[twocolumn]{emulateapj}

\usepackage{amsmath,amssymb}

\usepackage[colorlinks=true,allcolors=blue,urlcolor=magenta,bookmarks=false,pdfstartview=FitV]{hyperref}
\usepackage{color}
\usepackage{multirow}
\usepackage{appendix}
\usepackage[version=4]{mhchem}
\begin{document}

\title{CO Depletion in Protoplanetary Disks:\\A unified picture combining physical sequestration and chemical processing}

\author{Sebastiaan Krijt\altaffilmark{1,5,7}, Arthur D. Bosman\altaffilmark{2}, Ke Zhang\altaffilmark{2,5}, Kamber R. Schwarz\altaffilmark{3,6}, Fred J. Ciesla\altaffilmark{4,7} \& Edwin A. Bergin\altaffilmark{2}}
\email{skrijt@email.arizona.edu}

\altaffiltext{1}{Department of Astronomy/Steward Observatory, The University of Arizona, 933 North Cherry Avenue, Tucson, AZ 85721, USA}
\altaffiltext{2}{University of Michigan, Department of Astronomy, 1085 S. University, Ann Arbor, MI 48109, USA}
\altaffiltext{3}{Lunar and Planetary Laboratory, University of Arizona, 1629 E. University Blvd, Tucson, AZ 85721, USA}
\altaffiltext{4}{Department of the Geophysical Sciences, The University of Chicago, 5734 S. Ellis Avenue, Chicago, IL 60637, USA}
\altaffiltext{5}{Hubble Fellow}
\altaffiltext{6}{Sagan Fellow}
\altaffiltext{7}{EOS Team, NASA Nexus for Exoplanet System Science}

\begin{abstract} 
The gas-phase \ce{CO} abundance (relative to hydrogen) in protoplanetary disks decreases by up to 2 orders of magnitude from its ISM value ${\sim}10^{-4}$, even after accounting for freeze-out and photo-dissociation. Previous studies have shown that while local chemical processing of \ce{CO} and the sequestration of \ce{CO} ice on solids in the midplane can both contribute, neither of these processes appears capable of consistently reaching the observed depletion factors on the relevant timescale of $1{-}3\mathrm{~Myr}$. In this study, we model these processes simultaneously by including a compact chemical network (centered on carbon and oxygen) to 2D ($r+z$) simulations of the outer ($r>20\mathrm{~au}$) disk regions that include turbulent diffusion, pebble formation, and pebble dynamics. In general, we find that the $\ce{CO}/\ce{H2}$ abundance is a complex function of time and location. Focusing on \ce{CO} in the warm molecular layer, we find that only the most complete model (with chemistry and pebble evolution included) can reach depletion factors consistent with observations. In the absence of pressure traps, highly-efficient planetesimal formation, or high cosmic ray ionization rates, this model also predicts a resurgence of \ce{CO} vapor interior to the \ce{CO} snowline. We show the impact of physical and chemical processes on the elemental (C/O) and (C/H) ratios (in the gas and ice phases), discuss the use of CO as a disk mass tracer, and, finally, connect our predicted pebble ice compositions to those of pristine planetesimals as found in the Cold Classical Kuiper Belt and debris disks.
\end{abstract}


\section{Introduction}
The carbon-monoxide (\ce{CO}) molecule has played an important role in our understanding of protoplanetary disks. Being abundant, volatile, and having rotational transitions readily observable at millimeter wavelenghts, emission from gas-phase \ce{CO} and its isotopologues has been used to study, amongst other things, disk sizes \citep{andrews2012,ansdell2018,trapman2019,boyden2020}, gas masses \citep{ansdell2016,miotello2016}, temperature profiles \citep{zhang2017,dutrey2017,dullemond2020}, as well as detailed gas kinematics and their relation to turbulence \citep{flaherty2015,teague2016} and the presence of planets \citep[e.g.,][]{teague2018,pinte2019}. 

While the abundance of \ce{CO} (relative to hydrogen) is generally ${\sim}10^{-4}$ in the ISM and molecular clouds, a picture is beginning to emerge in which \ce{CO} becomes increasingly depleted in the warm ($T\gtrsim20\mathrm{~K}$) molecular layers of protoplanetary disks \citep{bergin2018}; regions in which \ce{CO} should be unaffected by freeze-out or photo-dissociation \citep{williams2014}.

First, many disks are surprisingly faint in \ce{CO} emission \citep{eisner2016}, which results in low estimates for disk gas masses \citep{ansdell2016,miotello2017,long2017}. Furthermore, in sources for which an independent measure of the hydrogen gas mass exists in the form of \ce{HD}, \ce{CO} abundances as low as $10^{-6}$ have been reported \citep{favre2013,bergin2013,mcclure2016}. Recently, \citet{zhang2020a}, by contrasting observations of multiple class 0/I and class II disks, found that \ce{CO} depletion factors of $10{-}100$ are common, and appear to be established on timescales of ${\sim}1{-}3\mathrm{~Myr}$ \citep[see also][]{bergner2020}.

Understanding the mechanism(s) behind the removal of \ce{CO} is important for several reasons. First, the accuracy of using \ce{CO} emission as a gas mass tracer relies on a firm understanding of what \ce{CO}/\ce{H2} should be. Second, \ce{CO} is an important carrier of carbon and oxygen, so if the underlying mechanism (whatever it is) will change the molecular and/or elemental abundances in the disk, affecting the compositions of (giant) planets forming there \citep[e.g.,][]{oberg2011,oberg2016,booth2017,eistrup2018,cridland2019,cridland2020}.

Several authors have studied chemical \ce{CO} processing as the source for the inferred depletion \citep[e.g.,][]{aikawa1996,bergin2014,reboussin2015,bosman2018b,schwarz2018,dodson-robinson2018,schwarz2019}. Using passive model disks (i.e., ignoring material transport), these authors generally find that chemical processing alone is inefficient on ${\sim}\mathrm{Myr}$ timescales in regions of the disk where \ce{CO} is not frozen out. Models with cosmic ray ionization rates $\zeta_\mathrm{CR} \gtrsim 10^{-17}\mathrm{~s^{-1}}$ can produce significant depletion factors in some cases \citep{bosman2018b}, but become inefficient in warmer disk regions, in disks with a significant of mass in pebble-size particles, and for low oxygen abundances \citep{schwarz2018}. Furthermore, it is unclear how appropriate ISM-level values for $\zeta_\mathrm{CR}$ are inside protoplanetary disks \citep{cleeves2015}.

Alternatively, others have considered physical sequestration of \ce{CO} in the cold ($\lesssim20\mathrm{K})$ outer disk midplane, often requiring of vertical transport with either grain growth \citep[e.g.,][]{kama2016,krijt2018}, or a vertically-varying diffusion coefficient \citep{xu2017}. A natural by-product of models that rely on dust coagulation and transport is that, in the absence of chemical processing, the inward radial drift of icy pebbles will increase the gas-phase CO abundance interior to the \ce{CO} snowline, in some cases by an order of magnitude \citep{piso2015,stammler2017,krijt2018}.

Recently, \citet{zhang2019} inferred radially resolved \ce{CO} depletion profiles for a handful of nearby disks (DM~Tau, TW~Hya, HD~163296, and IM~Lup), concluding that while the radial variations in the \ce{CO} abundance are similar to those predicted by \citet{krijt2018}, the observed depletion factors can be significantly higher than those found in the simulations, suggesting that \emph{both} ongoing chemical processing and dust evolution and transport play a role in setting the CO abundance. For HD~163296, the only case in which the region interior to the midplane \ce{CO} snowline could be resolve, evidence for an increased \ce{CO} abundance in the inner regions was found, and later confirmed using additional observations \citep{zhang2020b}.

While theoretical studies have started to combine physical and chemical evolution \citep{booth2017,bosman2018a,booth2019}, most of these works focused on the effect of inward radial transport on molecular abundances in the disk midplane, thus limiting their models to 1D without discussing the warm molecular layer.

We set out to study how processes associated with the earliest stages of planet/planetesimal formation (i.e., the formation of pebbles, their accumulation in the midplane, and their subsequent inward migration) interact with ongoing chemical processing of \ce{CO} (through either gas-phase or grain-surface reactions). Our main goal is to explore whether these processes, acting together, can explain the extreme \ce{CO} depletion factors, as well as their radial variations, that have been reported in the literature \citep[e.g.,][]{zhang2019,zhang2020a}, establishing a direct link between observational constraints probing the outer disk surface layers to physical and chemical processes taking place in the less accessible disk midplane.

\section{Methods}\label{sec:model}
We summarize here the various chemical and physical processes that are considered. 

\subsection{Disk structure}\label{sec:disk}
\begin{figure}
\centering
\includegraphics[clip=,width=1.\linewidth]{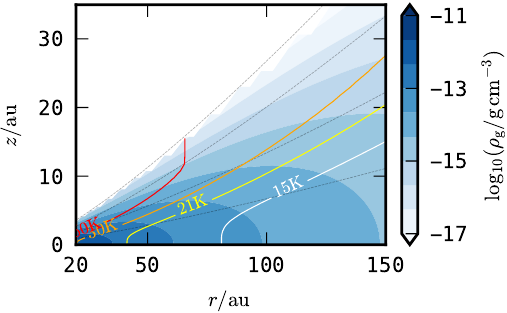}
\caption{Gas density and temperature structure used throughout this study (see Sect.~\ref{sec:disk}). Dashed lines indicate $z/H=\{1,2,3,4\}$.}.
\label{fig:tempdens}
\end{figure}

In this work, we focus on the region exterior to $r=20\mathrm{~au}$, as we are interested in the behavior around and outside the \ce{CO} snowline. When setting the disk gas 2D density and temperature structure (both assumed constant in time) we use the same approach as detailed in \citet[][Sect.~2.1]{krijt2018}, with minor changes to some of the global disk parameters. Specifically, focusing on a disk around a sun-like star, we have set the characteristic disk radius $r_c=80\mathrm{~au}$, disk mass $M_\mathrm{disk}/M_\star = 0.1$, and the midplane temperature $T_\mathrm{1au}=135\mathrm{~K}$, making the disk slightly smaller, twice as massive, and a little bit warmer compared to the one in \citet{krijt2018}. The gas density and temperature structure is illustrated in Fig.~\ref{fig:tempdens}.

At $t=0$, we assume all grains are $0.1\mathrm{~\mu m}$ in size, and present everywhere in the disk at a dust-to-gas mass ratio of $(\rho_\mathrm{d}/\rho_\mathrm{g}) = 10^{-2}$. For this setup, there is ${\approx}225~M_\oplus$ of dust (excluding ices) located beyond $r=20\mathrm{~au}$ at the beginning of the simulations.

\subsection{Chemistry}\label{sec:chemistry}
Our focus is on understanding five major carbon and oxygen carrying molecules: \ce{H2O}, \ce{CO}, \ce{CO2}, \ce{CH4}, and \ce{CH3OH}. These species have been chosen because together they represent the dominant carbon and oxygen-bearing species in the colder regions of the outer disk.

For these five molecules, we solve a set of (semi)analytical differential equations, which include the following \ce{CO} processing pathways:
\begin{enumerate}
    \item Successive hydrogenation of \ce{CO} ice, ultimately leading to the formation of \ce{CH3OH};
    \item \ce{CO} reacting with \ce{OH} on grain surfaces, leading to the formation of \ce{CO2} ice;
    \item Gas-phase \ce{CO} reacting with \ce{He+}, leading to the formation of \ce{CH4}.
\end{enumerate}
These pathways, and the details of how they are approximated here, are discussed more in \citet[][Sect.~2.3]{bosman2018b} and Appendix~\ref{sec:appendix_network}.

The initial abundances for the 5 species are listed in Table~\ref{tab:molecules}. These initial conditions are based on \citet[][Table~B.1]{bosman2018b}, and correspond to the majority of the carbon being in CO, while the oxygen is initially split between \ce{CO} and \ce{H2O} (see also Sect.~\ref{sec:appendix_network}). For a discussion of the validity of these assumptions see \citet[][Sect.~B.4.2]{bosman2018b}.

The timescales on which the chemical reactions take place is set by the cosmic ray ionization rate, which we set to $\zeta_\mathrm{CR}=10^{-17}\mathrm{~s^{-1}}$ in our standard model. The surface densities we are interested in do not exceed $100\mathrm{~g/cm^2}$, justifying the use of a constant $\zeta_\mathrm{CR}$ \citep{umebayashi1981,padovani2018}. We do not include X-rays as a source of ionization, the presence of which could potentially speed up \ce{CO} destruction in very low mass (${\leq}0.003M_\odot$) disks \citep{schwarz2018}.

\subsection{Freeze-out and desorption}
Desorption energies for the 5 major species are listed in Table~\ref{tab:molecules}. The balance between freeze-out and desorption on small dust grains is assumed to reach steady state for all five major molecular species during each timestep \citep[as in][]{bosman2018b}. This is a valid assumption in most regions of the disk, but can potentially lead to the models underpredicting the amount of gas-phase \ce{CO} in the cold outer disk regions, especially when these become depleted in dust \citep{semenov2006,krijt2018}. However, for the relatively weak turbulence strength used in this work, these effects will be minor. For pebble-size particles that drift through the midplane snowline, time-dependent sublimation is included, which will result in pebbles releasing \ce{CO} molecules over an extended radial region \citep[see also][]{piso2015}. 

For the initial abundances, the midplane \ce{CO} snowline (defined as the location where 50\% of the present \ce{CO} is frozen out) is located at $r\approx34\mathrm{~au}$, where the midplane temperature equals $T(r)\approx23\mathrm{K}$.

\subsection{Transport via turbulent diffusion}
The evolution of solids is handled in way that is very similar to the approach described in \citet{krijt2018}, with small grains being treated as a fluid on a grid, while the dynamics of pebble-size particles are followed using a swarm of representative tracer particles.

For the small grains, the ices present on small grains, as well as trace gas-phase vapor species, diffusive transport is included using the approach of \citet{krijt2018} \citep[see also][]{ciesla2009}. The strength of the turbulence is controlled via the dimensionless $\alpha$-viscosity parameter \citep{shakura1973}, which is assumed to be connected to the gas diffusivity via $D_\mathrm{g} = \alpha c_s H$. Motivated by \citet{flaherty2015,teague2016}, we set $\alpha=10^{-4}$, an order of magnitude lower than in the fiducial runs of \citet{krijt2018}.

\subsection{Pebble formation and evolution}\label{sec:dust_and_transport}
The formation and evolution of pebble-size particles is handled in way that is very similar to the approach described in \citet{krijt2018}. For increased flexibility and more control over the pebble behavior, however, we simplify the approach of \citet[][Sect.~3]{krijt2018} in two ways:
\begin{enumerate}
    \item We assume all pebbles that form in a single simulation have the same size $s_\mathrm{p}$, independent of their formation location.
    \item We approximate the dust-to-pebble conversion timescale as
    \begin{equation}
        \tau_\mathrm{d\rightarrow p} = \dfrac{a_1}{\Omega} \left(\dfrac{\rho_\mathrm{d}}{\rho_\mathrm{g}} \right)^{-1},
    \end{equation}
    with $\Omega$ the local Keplerian frequency and $a_1$ a constant.
\end{enumerate}
When pebbles form, they inherit the ice composition of the small grains present in that particular grid cell. The local abundance of small grains is lowered accordingly, and the effect of the decreasing small-dust-to-gas is included in the chemical network. 

Vertical settling, radial drift, and turbulent transport are included following \citep{krijt2018}. We include the sublimation of volatile species present on pebble-size particles if and when they move to disk regions with a higher temperature (i.e., as they drift inward), but otherwise pebbles are assumed to be chemically inert because of their small surface-to-mass ratio. In practice this means that small dust and pebbles can have quite different ice compositions, even when present at the same location. 

In the fiducial model, we will set $s_\mathrm{p}=1\mathrm{~mm}$ throughout the disk and, as in \citet{krijt2018}, imagine further growth is frustrated by a combination of bouncing and radial drift removing particles before they can gain more mass. We ignore collisional fragmentation, which is reasonable for the outer disk regions and the fairly low value of $\alpha=10^{-4}$ \citep{birnstiel2012,misener2019}. In more complex models, the exact maximum particle size is a function of location, local dust-to-gas ratio (for the drift limit), and particle composition, but a constant $s_\mathrm{p}=1\mathrm{~mm}$ suffices for our purposes and reproduces the typical pebble size outside ${\sim}20\mathrm{~au}$ in more complete models to within a factor of a few \citep[e.g.,][]{stammler2017,krijt2018}. For our disk model, pebbles of this size have Stokes numbers of $\mathrm{St}=4\times10^{-3}$, $0.02$, and $0.05$ in the midplane at $r=20$, $50$, and $100\mathrm{~au}$, respectively.

\subsection{Integration}
Starting with all the dust present as sub-micron grains and the initial abundanecs listed in Table~\ref{tab:molecules}, the 5 molecular tracers (in either vapor and/or ice form), the small dust, and pebble tracers are evolved forward in time. During every iteration:
\begin{enumerate}
    \item Gas-phase and grain-surface chemistry is advanced in each cell using the simplified scheme outlined in Appendix \ref{sec:appendix_network}.
    \item The new sublimation and freeze-out balance is calculated and the partitioning of volatile species between ice and vapor is updated in each cell.
    \item The exchange of small dust grains (carrying ices) and gas-phase species between neighbouring cells is calculated.
    \item The positions of the pebble tracer particles are updated, taking into account vertical settling, radial drift, and turbulent diffusion.
    \item The sublimation of ices carried by pebbles that have moved to warmer disk regions is calculated.
    \item In each cell, a fraction of the locally available dust mass is stochastically converted into pebble particles. The ice on newly-formed pebbles has the same composition as the dust in that cell.
\end{enumerate}
Steps $1{-}6$ are repeated until $t=3\mathrm{~Myr}$ have passed, and the timestep for this operative splitting routine is set by the shortest diffusion timescale across a single cell \citep[][Eq.~17]{krijt2018}. Similarly, we set $a_1=10$, which results in coagulation timescales similar to those found in 1D models (see Sect.~\ref{sec:NG06b}).

\begin{deluxetable}{l c c }[!b]
\centering
\tablecaption{Properties of molecular species that are followed.}
\tablewidth{0pt}
\tablehead{
Molecule  & \colhead{ Initial abundance} & \colhead{$E_\mathrm{bind}/\mathrm{K}$} 
}
\startdata
\ce{H_2O}             & $1.2\times10^{-4}$    & $5770$ \\
\ce{CO}             & $1.0\times10^{-4}$    & $855$ \\
\ce{CO_2}             & $1.0\times10^{-5}$    & $2990$ \\
\ce{CH3OH}             & $1.0\times10^{-6}$    & $4930$ \\
\ce{CH4}             & $1.0\times10^{-9}$    & $1200$ 
\enddata
\tablecomments{Desorption energies taken from \citet{bosman2018b}. Initial abundances in molecules per \ce{H}.}
\label{tab:molecules}    
\end{deluxetable}\vspace{1cm}

\section{Results}\label{sec:results}
In this Section we describe increasingly complex models by adding adding physical and chemical processes one by one, as summarized in Table~\ref{tab:modelruns}. Where possible, we compare the observed behavior to that seen in published studies that were similar in scope.

\begin{figure*}[t]
\centering
\includegraphics[clip=,width=1.\linewidth]{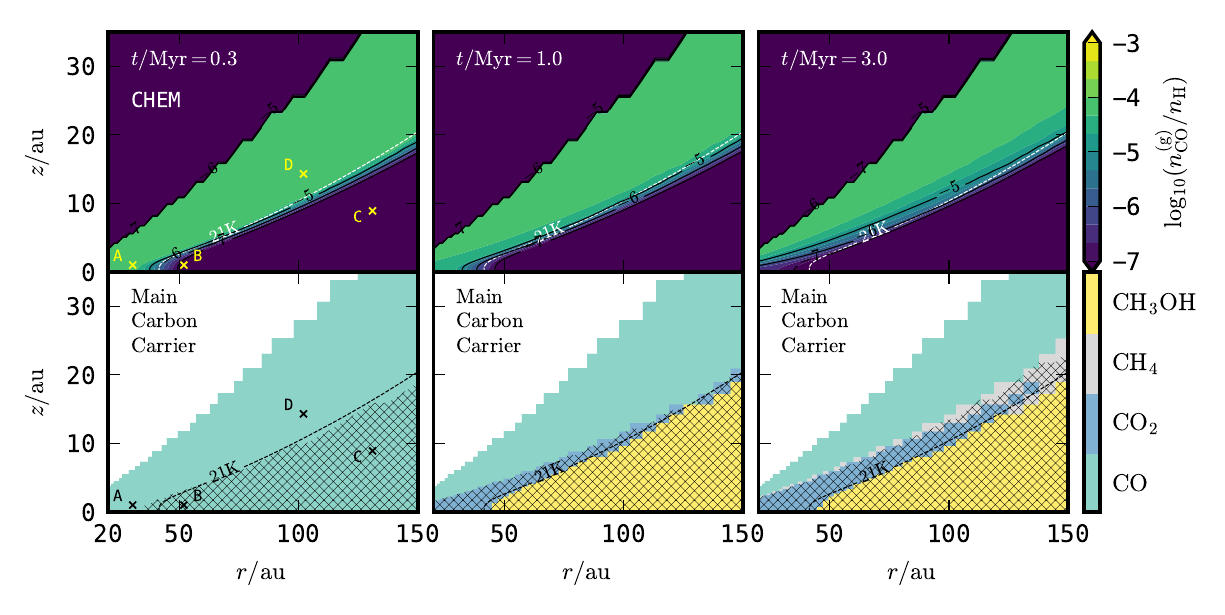}
\caption{Snapshots of the evolution of the model including chemical reaction only (\texttt{CHEM}, see Sect.~\ref{sec:NG01} and Table~\ref{tab:modelruns}). Top: Gas-phase \ce{CO} abundance as a function of location in the disk. Bottom: Main carbon carrier as a function of location in the disk. Shaded areas indicate the dominant carrier is present predominantly in solid form (i.e., frozen out on grain surfaces). The $T=21\mathrm{~K}$ contour roughly indicates the location where \ce{CO} transitions from being in the gas-phase (at higher temperatures) to being frozen out. Locations A, B, C, and D correspond to the temperature and density conditions shown in Fig.~\ref{fig:full_comparison} in more detail.}
\label{fig:NG01}
\end{figure*}

\subsection{Static chemistry only}\label{sec:NG01}
The starting point is a model in which we only include chemical processing and freeze-out/sublimation of molecular species. Dust coagulation and material exchange between adjacent grid cells is not included (see Table~\ref{tab:modelruns}). The results of this model are presented in Fig.~\ref{fig:NG01}, in which the following locations, correspond (roughly) to the temperature and density combinations shown in Fig.~\ref{fig:full_comparison}, have been highlighted:
\begin{itemize}
\item[A.]{Inside the CO snowline,}
\item[B.]{Outside the CO snowline,}
\item[C.]{Main freeze-out zone in the outer disk,}
\item[D.]{The warm molecular layer (WML) \citep[e.g.,][]{aikawa2002}.}
\end{itemize}
The observed behavior can be directly compared to the chemically more sophisticated, but similarly static studies \citep[e.g.,][and others]{walsh2010, henning2013,schwarz2018,bosman2018b}. Like \citet{schwarz2018} and \citet{bosman2018b}, we find that the processing of \ce{CO} in the warm molecular layer (e.g., location D) is inefficient on ${~}\mathrm{Myr}$ timescales. In colder regions, where \ce{CO} is either fully or partially frozen out on grain surfaces, more rapid processing can occur resulting typically in \ce{CO2} ice (around $T\sim20{-}25\mathrm{K}$, see location A) and \ce{CH3OH} or \ce{CH4} ice (where $T<21\mathrm{K}$, see locations B and C) becoming the dominant carbon carrier \citep[see also Sect.~\ref{sec:chemistry} and][Figs.~5 and 7]{bosman2018b}. Similar findings across a variety of disk temperature and density profiles led \citet{zhang2019} to conclude that chemical processing alone is not an efficient way of removing the warm \ce{CO} gas that dominates the emission readily probed by ALMA.

\begin{figure*}
\centering
\includegraphics[clip=,width=1.\linewidth]{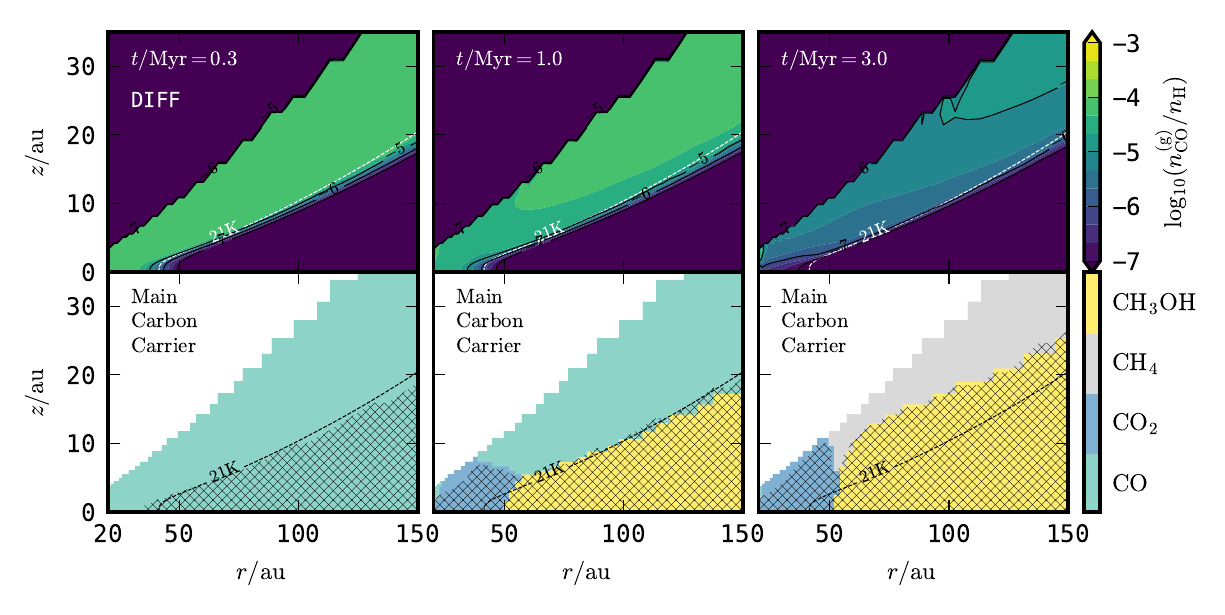}
\caption{Snapshots of the evolution of the model including chemical reactions and diffusive transport of vapor and small solids (\texttt{DIFF}, see Sect.~\ref{sec:NG02} and Table~\ref{tab:modelruns}).}
\label{fig:NG02}
\end{figure*}

\subsection{Chemistry and turbulent diffusion}\label{sec:NG02}
The next step is to include material transport through turbulent diffusion. In the absence of dust coagulation, and thus large variations in the dust-to-gas ratio, abundance gradients are set mainly by spatial variations in temperature, gas density, and the radiation field. Generally, diffusion will act to smear out such spatial gradients in the concentrations of gas-phase and ice-phase species \citep[e.g.,][]{willacy2006,ciesla2009,semenov2011}.

The timescale for vertical mixing to act on a scale $\Delta x$ can be estimated as $t_z = (\Delta x)^2/D_\mathrm{g}$. Focusing on vertical transport. For $\Delta x=H$ and assuming a diffusion coefficient based on a turbulence strength described by $\alpha=10^{-4}$ (Sect.~\ref{sec:disk}), we obtain $t_z \approx 0.25$ and $1.6~\mathrm{Myr}$ at $30$ and $100\mathrm{~au}$, respectively. These timescales are shorter than or comparable to the chemical processing in the disk midplane (Figs.~\ref{fig:NG01} and \ref{fig:full_comparison}), suggesting turbulent transport and chemical processing of \ce{CO} are intertwined and the \ce{CO} abundance is sensitive to transport processes.

The results of the chemistry and diffusion (\texttt{DIFF}) model are shown in Fig.~\ref{fig:NG02}. Comparing the behavior to the static \texttt{CHEM} model (Fig.~\ref{fig:NG01}), the effect of turbulent transport is indeed evident. First, the ice species produced near the midplane are effectively transported to the disk's upper layers, resulting in \ce{CH3OH} and \ce{CH4} being present at high $z/H$ after $3~\mathrm{Myr}$. Second, the narrow band of \ce{CO2} seen hugging the $T\sim20{-}25~\mathrm{K}$ region in Fig.~\ref{fig:NG01} has disappeared, with \ce{CO2} ice only being the dominant carbon carrier for $r<50\mathrm{~au}$. In this region, however, fast vertical mixing has resulted in \ce{CO2} being dominant throughout the entire vertical column. Qualitatively similar behavior has been found in earlier works by \citet{semenov2011,furuya2014}.
 
Finally, Fig.~\ref{fig:NG02} reveals a decrease in the gas-phase \ce{CO} abundance in the WML of about an order of magnitude after $3~\mathrm{Myr}$. This can be understood in terms of vertical mixing: While local processing of \ce{CO} in the WML is not efficient (see Fig.~\ref{fig:NG01}), the addition of diffusion allows gas-phase \ce{CO} molecules to be cycled through the colder midplane, where (1) they rapidly stick to grain surfaces, and (2) can be processed to form less volatile species such as \ce{CH3OH}. These products, and the portion of \ce{CO} ice that escaped processing, will eventually return to the WML, but overall the process is asymmetric and results in the loss of gas-phase \ce{CO} from the upper regions on ${\sim}\mathrm{Myr}$ timescales. 

In the seminal work of \citet{semenov2011}, the depletion of gas-phase \ce{CO} in the outer disk was found to be inefficient \citep[see also][]{semenov2006}, its column density changing by no more than a factor ${\sim}3{-}5$ over $5\mathrm{~Myr}$ when turbulent diffusion was included. While a direct comparison is difficult because of several differences in the set-up (disk density and temperature structure, etc.), we attribute this difference to two reasons: (1) \citet{semenov2011} did not include \ce{H} and \ce{H2} tunneling. This results in a slower conversion of \ce{CO} to \ce{CH3OH} on the grain surface, while making it easier to re-form \ce{CO} as the incorporation of \ce{O} into \ce{H2O} is also suppressed \citep[see also][]{bosman2018b}; and (2) \citet{semenov2011} used a ratio of diffusion to binding energy $E_\mathrm{diff}/E_\mathrm{bind}=0.77$, substantially larger than our value of 0.3, also reducing the efficiency of grain surface chemistry.

\begin{figure*}[t]
\centering
\includegraphics[clip=,width=1.\linewidth]{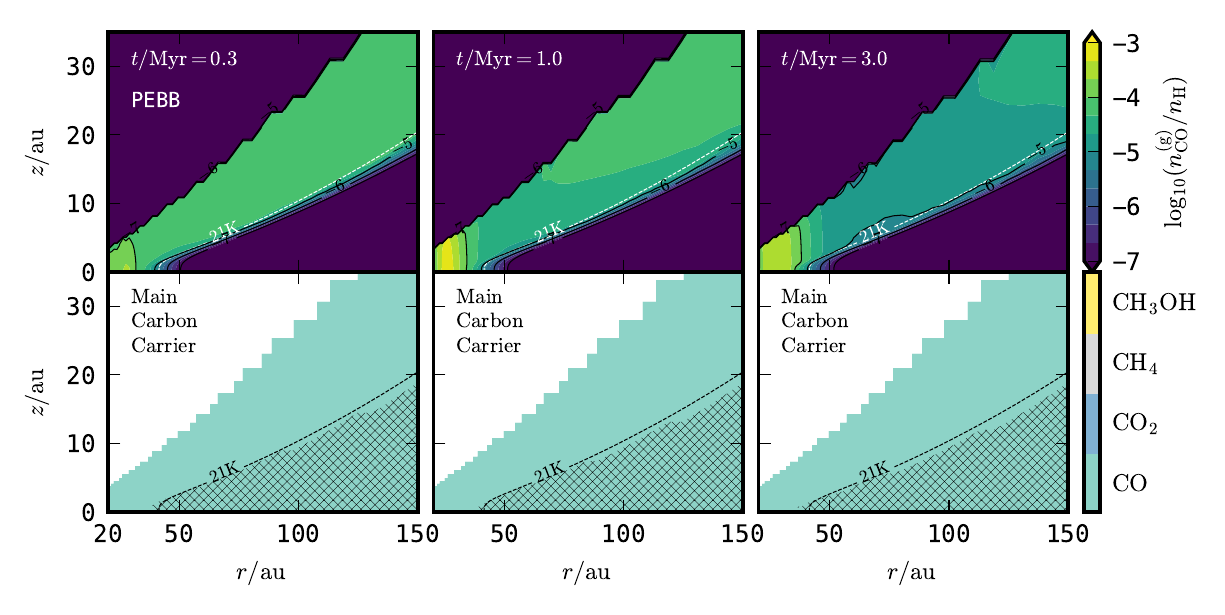}
\caption{Snapshots of the evolution of the model including only dust coagulation and pebble migration (\texttt{PEBB}, see Sect.~\ref{sec:NG06b} and Table~\ref{tab:modelruns}).}
\label{fig:NG06b}
\end{figure*}

\begin{deluxetable}{l c c c c }[!b]
\centering
\tablecaption{Processes included/excluded in the models of Sect.~\ref{sec:results}.}
\tablewidth{0pt}
\tablehead{
Model $\rightarrow$ & \colhead{ \texttt{CHEM} } & \colhead{\texttt{DIFF}}  & \colhead{\texttt{PEBB}}  & \colhead{\texttt{FULL}}
}
\startdata
Chemical processing             & \checkmark    & \checkmark 	&$\times$   	& \checkmark	\\
Freeze-out/sublimation 			& \checkmark 	& \checkmark 	& \checkmark 	& \checkmark	\\
Vapor diffusion 				& $\times$ 	    & \checkmark 	& \checkmark 	& \checkmark	\\
Dust and ice dynamics 			& $\times$ 	    & \checkmark 	&\checkmark 	& \checkmark	\\ 
Pebble formation \& dynamics    & $\times$ 	    & $\times$   	&\checkmark 	& \checkmark	\\ 
Introduced in Figure            & \ref{fig:NG01}&\ref{fig:NG02} &\ref{fig:NG06b}&\ref{fig:NG07b}
\enddata
\tablecomments{These models all employ $s_\mathrm{p}=1~\mathrm{mm}$, $a_1=10$, and $\zeta_\mathrm{CR}=10^{-17}\mathrm{~s^{-1}}$. Variations on the \texttt{FULL} model are presented in Sect.~\ref{sec:discussion}.}
\label{tab:modelruns}    
\end{deluxetable}\vspace{1cm}

\subsection{Pebble formation and dynamics only}\label{sec:NG06b}
For completeness, we include here a model with dust evolution and transport only, without taking into account chemical processing (\texttt{PEBB}, Fig.~\ref{fig:NG06b}). This set-up is essentially identical to the one used in \citet{krijt2018}, with the caveat that we have made several minor changes to the way the pebble size and coagulation timescale are calculated (e.g., Sect.~\ref{sec:dust_and_transport}). 

As the result of pebble formation and vertical settling, the local dust-to-gas ratio decreases in the disk surface layer, and increases in the midplane \citep[e.g.,][Fig.~6]{krijt2018}. At the same time, the inward drift of pebble-size particles results in a radial mass flux of solids throughout the disk. Typically, at several 10s of au, this flux builds up rapidly, peaks around $10^5\mathrm{~yr}$, and then decreases slowly on ${\sim}10^6\mathrm{~yr}$ timescales \citep[][Fig.~12]{krijt2018}, although the detailed behavior varies with location and depends on the (initial) disk properties and details of the dust coagulation process \citep[e.g.,][]{birnstiel2012,lenz2019}. 

Figure~\ref{fig:pebble_masses} shows the time evolution of the total mass of various pebble reservoirs present in the \texttt{PEBB} simulation as well as the cumulative pebble mass that has drifted through the inner boundary. Over the course of the \texttt{PEBB} simulation, ${\sim}220 M_\oplus$ worth of pebbles can be seen to drift inward of $r=20\mathrm{~au}$, with about $10\%$ of them doing so in the first $0.2\mathrm{~Myr}$. Pebble fluxes of similar magnitudes play an important role in modern planet formation theories, providing a means for planetary embryos to accumulate mass quickly through pebble accretion \citep{johansen2017,ormel2017}. For example, roughly ${\sim}150M_\oplus$ of pebbles are required to grow the cores of the solar system's giant planets within $3\mathrm{~Myr}$ \citep{lambrechts2014}.

The consequences of pebble formation and dynamics on gas-phase \ce{CO} abundances have been discussed in \citet{krijt2018} and we recover the same behavior: (1) the \ce{CO} abundance in the WML drops by a factor of a few as ices are sequestered on pebble-size particles in the midplane; and (2) the sublimation of rapidly-drifting and ice-covered pebbles results in a plume of gas-phase \ce{CO} interior to the \ce{CO} ice-line. The appearance of this plume will vary with time as the radial pebble flux evolves (see Fig.~\ref{fig:pebble_masses}), while its shape and peak \ce{CO} abundance depend sensitively on the strength of the turbulence and its relation to the diffusion coefficient \citep[][Figs.~7 and 8]{stammler2017}.

\begin{figure*}
\centering
\includegraphics[clip=,width=1.\linewidth]{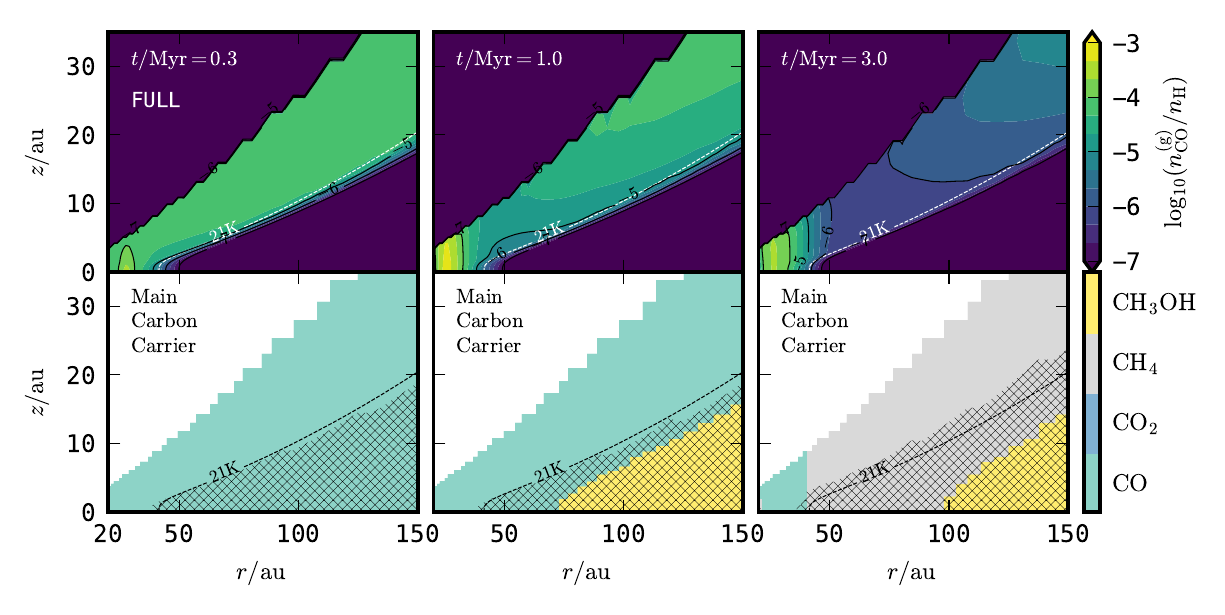}
\caption{Snapshots of the most complete model presented here, which includes chemical reactions, dust coagulation, as well as material transport (\texttt{FULL}, see Sect.~\ref{sec:NG07b} and Table~\ref{tab:modelruns}).}
\label{fig:NG07b}
\end{figure*}

\subsection{Chemistry, pebble evolution, and diffusive transport}\label{sec:NG07b}
Finally, we present the results of the \texttt{FULL} model including chemistry, diffusive transport, coupled with dust coagulation and pebble dynamics in Fig.~\ref{fig:NG07b}.

\begin{figure}
\centering
\includegraphics[clip=,width=1.\linewidth]{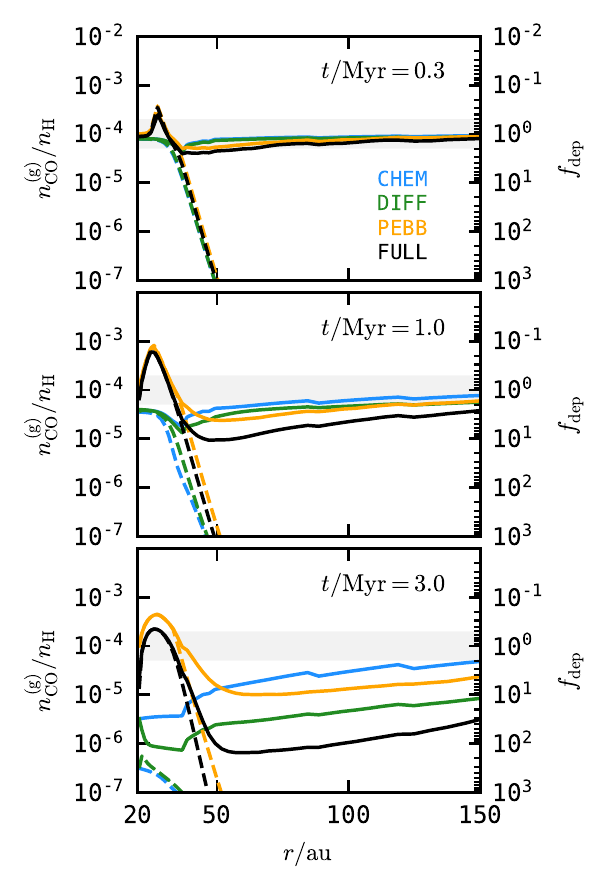}
\caption{Radial profiles of the gas-phase \ce{CO} abundance for the models presented in Figs.~\ref{fig:NG01}{-}\ref{fig:NG07b} as measured in the midplane (dashed curves) and the surface layer where $T>21\mathrm{~K}$ (solid). The depletion factor $f_\mathrm{dep}$ indicates the fractional change in the gas-phase \ce{CO} abundance relative to the initial value of $10^{-4}$, and is most meaningful for the surface layer curves and the midplane region interior to the \ce{CO} ice-line. The horizontal grey band shows values within a factor of 2 of the initial CO abundance.}
\label{fig:profiles}
\end{figure}

\begin{figure}
\centering
\includegraphics[clip=,width=1.\linewidth]{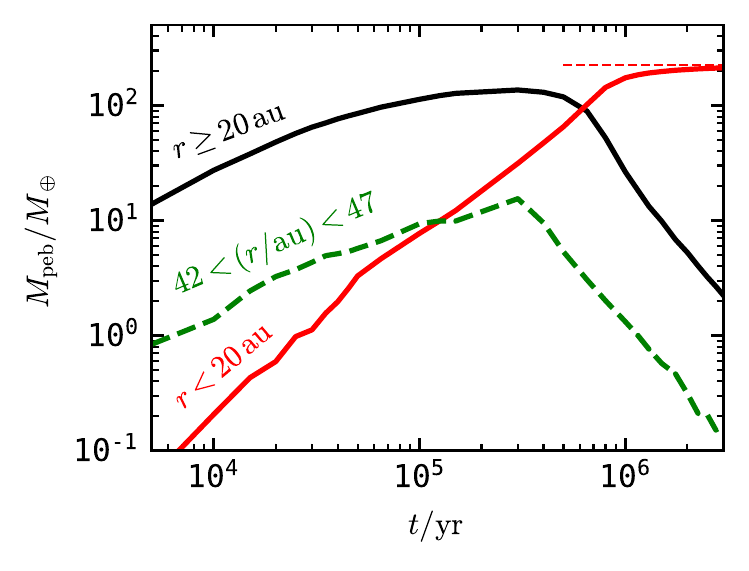}
\caption{Total mass of pebbles formed in the \texttt{PEBB} and \texttt{FULL} models (Figs.~\ref{fig:NG06b} and \ref{fig:NG07b}) that: are still present in the simulation domain (solid black); are currently located between $42{-}47\mathrm{~au}$ (dashed green); or have drifted interior to $r=20\mathrm{~au}$ region (solid red). The thin red dashed line shows the initial solid content of the region ${>}20\mathrm{~au}$.}
\label{fig:pebble_masses}
\end{figure}

Since the processing of \ce{CO} takes place on $\mathrm{Myr}$ timescales, the picture at early times (e.g., at $0.3\mathrm{~Myr}$) is comparable to the one presented in Fig.~\ref{fig:NG06b}, with a similar plume of gas-phase \ce{CO} developing centered around $r\sim30\mathrm{~au}$, and some \ce{CO} disappearing from the WML around $r\sim50\mathrm{~au}$ as it becomes sequestered on pebbles in the midplane. As in the \texttt{PEBB} model without chemistry, the plume inside the \ce{CO} ice-line persists for $3\mathrm{~Myr}$, but the magnitude at the end of the simulation is significantly reduced. This can be seen clearly in Fig.~\ref{fig:profiles}, where we compare radial profiles of the gas-phase \ce{CO} abundance (in the midplane and in the WML) for the 4 models described in Sects.~\ref{sec:NG01}{-}\ref{sec:NG07b}. There are two reasons for this reduction. First, \ce{CO} can be processed locally, mainly forming \ce{CO2} at these pressures and temperatures (see Fig.~\ref{fig:NG01}). Second, as a result of chemical processing on grain surfaces, pebbles that form and arrive at the \ce{CO} ice-line late in the simulation will carry a significant fraction of their carbon in the form of \ce{CH3OH} and \ce{CH4} rather than \ce{CO} (we come back to this point in Sect.~\ref{sec:KBOs}). These species have their ice-lines further in, and will not be released to the gas-phase as these pebbles cross the \ce{CO} ice-line. 

In the region inside and around the \ce{CO} ice-line, i.e., for $r<50\mathrm{~au}$, the inward migration and evaporation of \ce{CO}-ice-coated pebbles results in gas-phase \ce{CO} always being the dominant carbon carrier. The situation in the outer disk, however, especially for $t\gtrsim 1~\mathrm{Myr}$, resembles more closely the picture presented in Fig.~\ref{fig:NG02}, with \ce{CH4} and \ce{CH3OH} becoming the dominant carbon carriers. Focusing on the disk surface layer exterior to $r\sim50\mathrm{~au}$, we see that the two \ce{CO} removal mechanisms encountered earlier (sequestration on pebble surfaces, and turbulent transport followed by chemical processing in the midplane) seem to exacerbate the depletion of gas-phase \ce{CO}, with local abundances dropping by up to 2 orders of magnitude over the modelled $3\mathrm{~Myr}$. Such depletion factors are reached exclusively in the \texttt{FULL} model.

\section{Discussion}\label{sec:discussion}

\subsection{Pebble drift vs. chemical processing}\label{sec:substructures}
\begin{figure*}
\centering
\includegraphics[clip=,width=.45\linewidth]{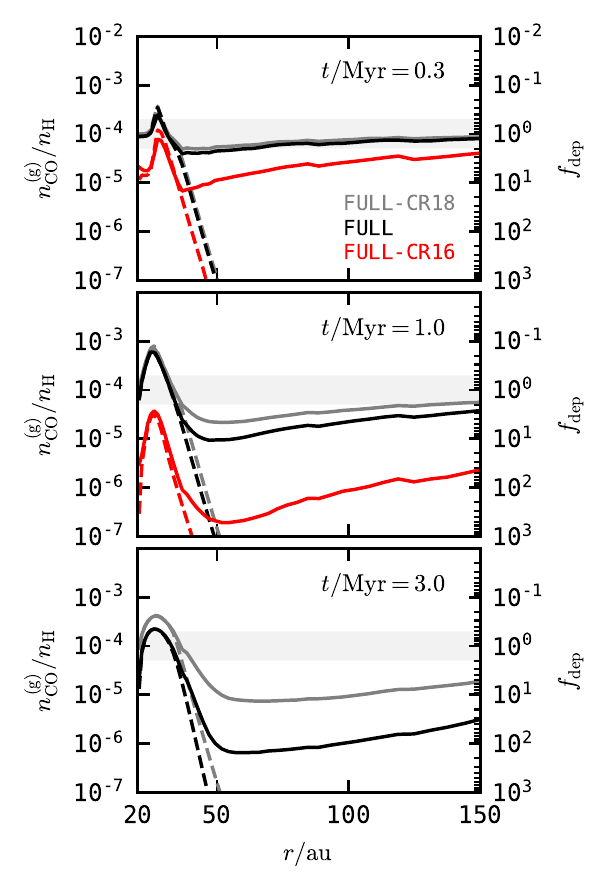}
\includegraphics[clip=,width=.45\linewidth]{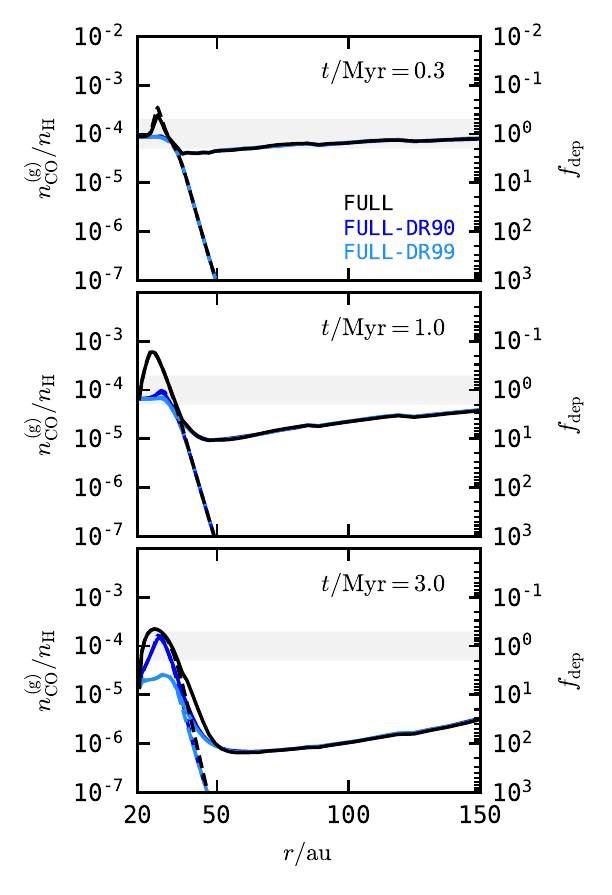}
\caption{Impact on radial CO depletion profiles for varying cosmic ray ionisation rates $\zeta_\mathrm{CR}$ (left) and pebble drift efficiencies (right). These models are discussed in Sect.~\ref{sec:discussion}.}
\label{fig:profiles_2}
\end{figure*}

In the framework presented here, there is nothing stopping pebbles from drifting inward in the models that include dust coagulation (Figs.~\ref{fig:NG06b} and \ref{fig:NG07b}). Thus, the only way to prevent a resurgence of gas-phase CO just inside the snowline from occurring is to have chemistry outpace pebble drift, resulting in pebbles carrying carbon in a less volatile form (i.e., in \ce{CH4} and \ce{CH3OH} in our calculations). This is illustrated in the left panels of Fig.~\ref{fig:profiles_2}, where we have varied $\zeta_\mathrm{CR}$ from $10^{-18}\mathrm{~s^{-1}}$ (\texttt{FULL-CR18}), $10^{-17}\mathrm{~s^{-1}}$ (\texttt{FULL}, the fiducial model), and $10^{-16}\mathrm{~s^{-1}}$ (\texttt{FULL-CR16}). We note that the overall ionization structure of disk systems as a function of time is highly uncertain \citep[e.g.,][]{cleeves2013,cleeves2015,padovani2018} and the values used here represent a reasonable range for exploration. The pebble behavior has not been altered in these runs, so the flux of solids crossing the \ce{CO} snowline is identical\footnote{Not counting variations in their ice content.} to the one depicted in Fig.~\ref{fig:pebble_masses}. In the high cosmic ray model, the region with enhanced \ce{CO} abundance around $20{-}40\mathrm{~au}$ persists for less than a million years, with gas-phase \ce{CO} being virtually absent after $3\mathrm{~Myr}$. The disappearance of the peak has 2 causes; (1) gas-phase \ce{CO} is processed locally, mostly converted to \ce{CO2} (see Sect.~\ref{sec:NG01}), and (2) chemical processing in the outer disk midplane means that pebbles arriving at times $t\gtrsim10^5\mathrm{~yr}$ contain very little \ce{CO} ice (see also Sect.~\ref{sec:KBOs} and Fig.~\ref{fig:pebbles45}).

Alternatively, the balance between chemical processing and radial drift can shift when pebble migration becomes less efficient or is completely halted. For example, the formation of larger bodies (i.e., planetesimals, see Sect.~\ref{sec:KBOs}) or the trapping of pebbles in radial pressure bumps should reduce the radial pebble flux \citep[e.g.,][]{pinilla2020}. In disks where massive planets have already formed, deep gaps in the gas surface density profiles can trap pebble-size particles outside the planet's orbit with a near 100\% efficiency \citep{pinilla2012,bae2019}. Indeed, pebble accumulations observed with ALMA (e.g., by the DSHARP survey) are regularly located exterior to the inferred \ce{CO} ice-line \citep{huang2018,long2018,andrews2020}, clearly demonstrating the viability of trapping pebbles and the ices they carry in the outer regions of protoplanetary disks. 

We illustrate the impact of these processes in the right-hand panels of Fig.~\ref{fig:profiles_2}, where we compare the fiducial model to a case where the drift efficiency is reduced by 90\% (\texttt{FULL-DR90}), and 99\% (\texttt{FULL-DR99}), respectively. It is evident that the \ce{CO} depletion pattern in the outer disk WML is identical, as this is set mainly by vertical transport. The decreasing importance of radial motions for pebbles has a clear impact on the \ce{CO} peak interior to the snowline. For the \texttt{FULL-DR90} model, a small peak still develops, but it takes several $\mathrm{Myr}$ to do so. In the \texttt{FULL-DR99} model, where pebble drift is essentially stopped (as would be the case when pebbles are converted to planetesimals very efficiently at all radial locations), even the region interior to the \ce{CO} snowline becomes depleted in \ce{CO}.

\subsection{Comparison to \citet{zhang2019,zhang2020a}}
Recently, \citet{zhang2019} derived radial profiles of the (gas-phase) \ce{CO} depletion factor\footnote{The definition of the depletion factor is slightly different in this work, but in practice they are identical.} by combining spatially resolved \ce{CO} isotopologue emission with SED modeling and thermo-chemical modeling for a handful of nearby disks (specifically, DM~Tau, TW~Hya, HD~163296, and IM~Lup). While these disks vary in age, size, and temperature structure, we can qualitatively compare their findings to the results of our models. 

Focusing on the WML region (solid curves) located beyond the \ce{CO} ice-line, Fig.~\ref{fig:profiles} indicates that depletion factors between ${\sim}10{-}100$, as inferred for IM~Lup and TW~Hya, are exclusively reached in the \texttt{FULL} model, which combined chemical processing with turbulent mixing and the sequestration of \ce{CO} ice in the disk midplane. Even with these processes combined, however, several million years of evolution were needed to reach such severe \ce{CO} depletion, while IM~Lup is believed to be younger, ${\sim}1\mathrm{~Myr}$ \citep{zhang2020b}. Possible reasons for more rapid \ce{CO} removal could include higher ionization rate (Fig.~\ref{fig:profiles_2}), higher turbulence in the disk surface layers \citep{xu2017,krijt2018}, or a colder disk in which the \ce{CO} surface snowline lies further away from the midplane \citep[][Fig.~4]{zhang2019}.

For HD~163296, the radial profile derived by \citet{zhang2019} included the region just interior to the \ce{CO} snowline, suggesting the presence of a plume of \ce{CO} vapor, as predicted in our models that included pebble drift (Sects.~\ref{sec:NG06b} and \ref{sec:NG07b}) and earlier works by \citet{stammler2017,krijt2018}. This local CO plume was further confirmed by the modeling of $^{13}$C$^{18}$O (2-1) line spectrum of HD 163296 \citep{zhang2020a}, who found that a radial pebble flux of ${\sim}15{-}60~M_\oplus/\mathrm{Myr}$ is needed to reproduce the elevated C/H ratio interior to the \ce{CO} snowline. Comparable fluxes\footnote{Note however that the number obtained in \citet{zhang2020a} corresponds to the flux measured at $70\mathrm{~au}$.} are seen in our models that include pebble evolution (see Sect.~\ref{sec:NG06b}). Observed substructure in the dust continuum emission of HD~163296 \citep[e.g.,][]{isella2018} might suggest efficient trapping of pebbles; however, \citet{rosotti2020} argue that even the largest grains currently present in the disk are relatively well coupled to the gas, finding $(\mathrm{St}/\alpha)\sim20$ for the two rings firmly outside the \ce{CO} snowline\footnote{ For comparison, the pebbles in the \texttt{PEBB} and \texttt{FULL} models have $(\mathrm{St}/\alpha)\sim200$ at $r=100\mathrm{~au}$.}. Such observational constraints on the radial pebble flux are very valuable because this quantity plays a key role in shaping the final masses and orbital architectures of both gas giant and more terrestrial planets \citep[e.g.,][]{johansen2017,lambrechts2019}. Lastly, the limited CO depletion in the outer disk for the relatively old HD~163296 are likely a result of the disk being fairly warm \citep{zhang2019,dullemond2020}, rendering \ce{CO} processing less efficient \citep{bosman2018b, schwarz2018}.

For TW~Hya, \ce{CO} isotopologue emmision reveals a significant amount of gas-phase \ce{CO} between $5{-}20\mathrm{~au}$, with abundances significantly higher than in the outer WML, but still depleted relative to ISM levels \citep{schwarz2016,zhang2017,zhang2019}. Recent studies by \citet{bosman2019} and \citet{mcclure2019} also argue \ce{CO} release inside $20\mathrm{~au}$ is minimal, suggesting carbon is either transported inward in a different form, or that pebble migration has ceased alltoghether (see Sect.~\ref{sec:substructures}). However, as discussed in \citet[][Sect.~6.3]{kama2016}, the C/Si ratio of gas accreting onto TW~Hya is an order of magnitude above that of typical T~Tauri stars, suggesting that some carbon makes it to the inner disk even as refractory elements like Si are locked up in planetesimals or even larger bodies.

We note that while the disk used in this work was quite massive (see Sect.~\ref{sec:disk}), we expect \ce{CO} depletion to proceed similarly in lower mass disks as the relevant timescales do not depend on disk gas mass directly, but rather on the dust-to-gas ratio (for pebble formation) and the turbulence strength (for vertical mixing). Nonetheless, the detailed temperature structure and ionization environment will set the timescale for chemical processing as well as the locations of snowlines. As such, in depth comparisons to observed protoplanetary disks warrant dedicated models for each individual object \citep[e.g.,][]{zhang2019}.

\begin{figure*}
\centering
\includegraphics[clip=,width=.95\linewidth]{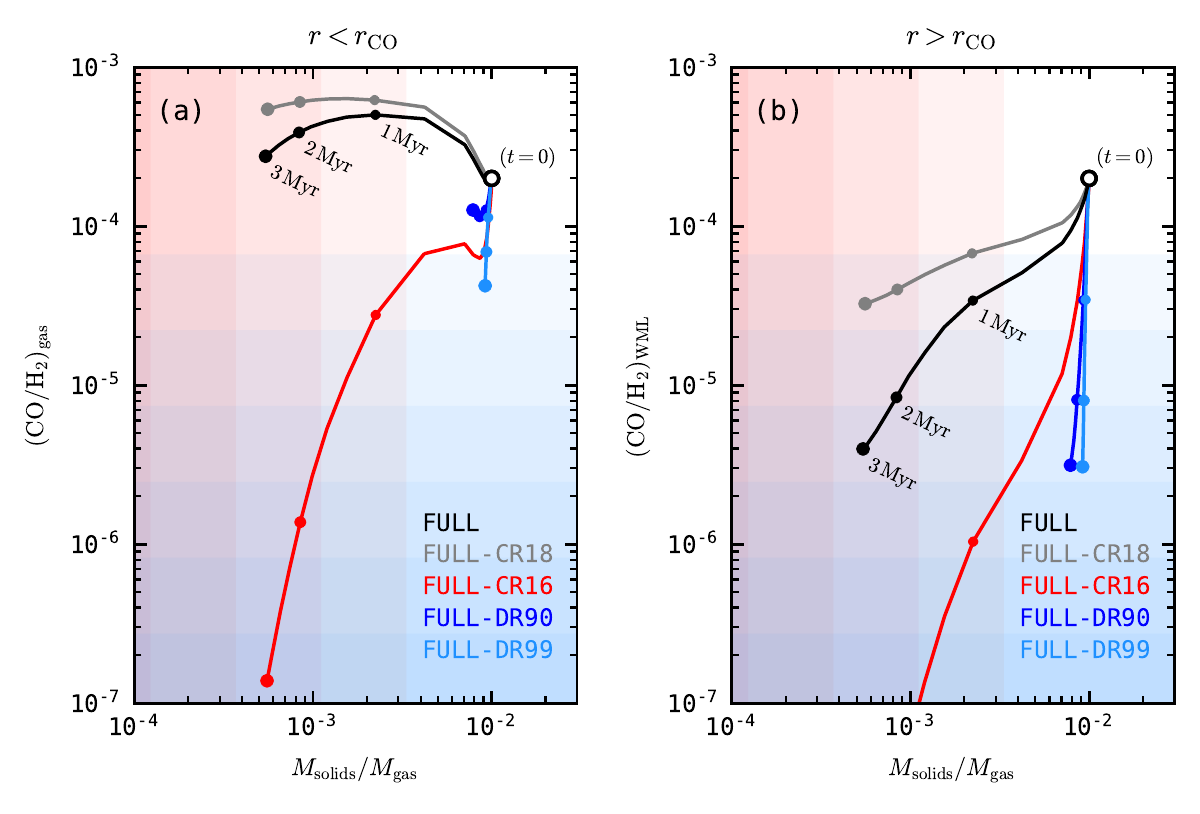}
\caption{Evolution of the (mass-weighted) \ce{CO/H2} in the gas inside the midplane CO snowline (left) and in the warm molecular layer (WML) outside the CO midplane snowline (right), compared to the total mass of solids (dust + pebbles) present exterior to $r=20\mathrm{~au}$ (both panels). Background colors represent factors of $3{\times}$ reductions in the CO abundance (blue) or solids-to-gas ratio (red).}
\label{fig:disk_mass}
\end{figure*}

\subsection{Implications for disk mass estimates}\label{sec:disk_mass}
In the absence of a direct gas mass tracer, gas-phase \ce{CO} (isotopologue) line fluxes and dust continuum emission originating from the outer disk are frequently used to estimate disk gas masses. A common approach is then to assume ISM-like values for $\ce{CO/H2}\approx10^{-4}$ and/or $M_\mathrm{dust}/M_\mathrm{gas}\approx10^{-2}$ \citep[e.g.,][and many others]{williams2014,miotello2016,ansdell2016,pascucci2016}. Uncertainties in these indirect approaches have recently been discussed in e.g., \citet{bergin2018,kama2020}. Indeed, using an alternative method of looking at disk dusk lines for several bright protoplanetary disks, \citet{powell2019} obtain typical dust-to-gas ratios of ${\sim}10^{-3}$ in the outer disk regions, and total gas disk masses between ${\sim}3{-}100{\times}$ higher than those derived from \ce{CO}.

In our simulations, both the amounts of warm \ce{CO}, dust, and pebbles vary significantly with time as a result of chemical processing and dust evolution. We illustrate the effects these processes can have on disk mass estimates in Fig.~\ref{fig:disk_mass}, which shows the time evolution of the mass-weighted \ce{CO/H2} and solids-to-gas conversion factors for the various models shown in Fig.~\ref{fig:profiles_2}. The left panel shows warm \ce{CO} interior to the midplane \ce{CO} snowline, while the panel on the right illustrates the warm molecular layer in the outer disk. In both panels, horizontal axis represents the total (i.e., dust+pebble) solid-to-gas mass ratio integrated over the entire disk. Background colors represent factors of $3{\times}$ reductions in the CO abundance (blue) or solids-to-gas ratio (red).

The trajectories depicted in Fig.~\ref{fig:disk_mass} illustrate that the accuracy of using either \ce{CO} emission (while assuming an ISM-like abundance) or dust emission (while assuming an ISM-like dust-to-gas ratio) as a total disk mass tracer depends sensitively on the details of the ongoing chemical and dust evolution. For example, in the \texttt{FULL-DR90} and \texttt{FULL-DR99} models, the reduced drift efficiency mean the the solid mass is hardly changing, while the warm \ce{CO} in the outer disk drops by a factor ${\sim}10^2$ over the modelled $3\mathrm{~Myr}$. In disks evolving in such a fashion, the dust mass would be a more reliable gas mass tracer, as found to be the case in the Lupus star forming region by \citet{manara2016}. Conversely, in the \texttt{FULL-CR18} model (see Sect.~\ref{sec:substructures}), slow processing of \ce{CO} and fairly rapid dust drift result in \ce{CO} emission from the outer regions being a more reliable gas mass tracer (especially at times ${\lesssim}\mathrm{Myr}$), even if the \ce{CO/H2} conversion factor has to be adjusted by a factor of a few. For the canonical \texttt{FULL} model, both the total solid and warm \ce{CO} mass in the outer disk are reduced by over an order of magnitude in $3~\mathrm{Myr}$. 

Generally, however, both the warm CO (at least in the outer disk) and total solid masses are decreasing functions of time, implying that studies looking for evolutionary trends in gas disk masses by comparing star forming regions of different ages have to be careful when using \ce{CO/H2} or dust-to-gas conversion factors that are constant in time, especially if a significant fraction of the disk mass is located beyond the \ce{CO} snowline.

\subsection{Elemental ratios and hydrocarbon production}\label{sec:hydrocarbons}
The processes described described in Sect.~\ref{sec:results} will alter key elemental ratios in different ways. Recognizing that the effect of radial drift on gas and solid elemental abundances in the midplane has been discussed in detail by multiple authors \citep[e.g.,][]{cuzzi2004,ciesla2006, estrada2016, oberg2016, booth2017, booth2019}, we focus here primarily on the warm molecular layer.

\subsubsection{Gas-phase C/H and C/O}
\begin{figure*}
\centering
\includegraphics[clip=,width=1.\linewidth]{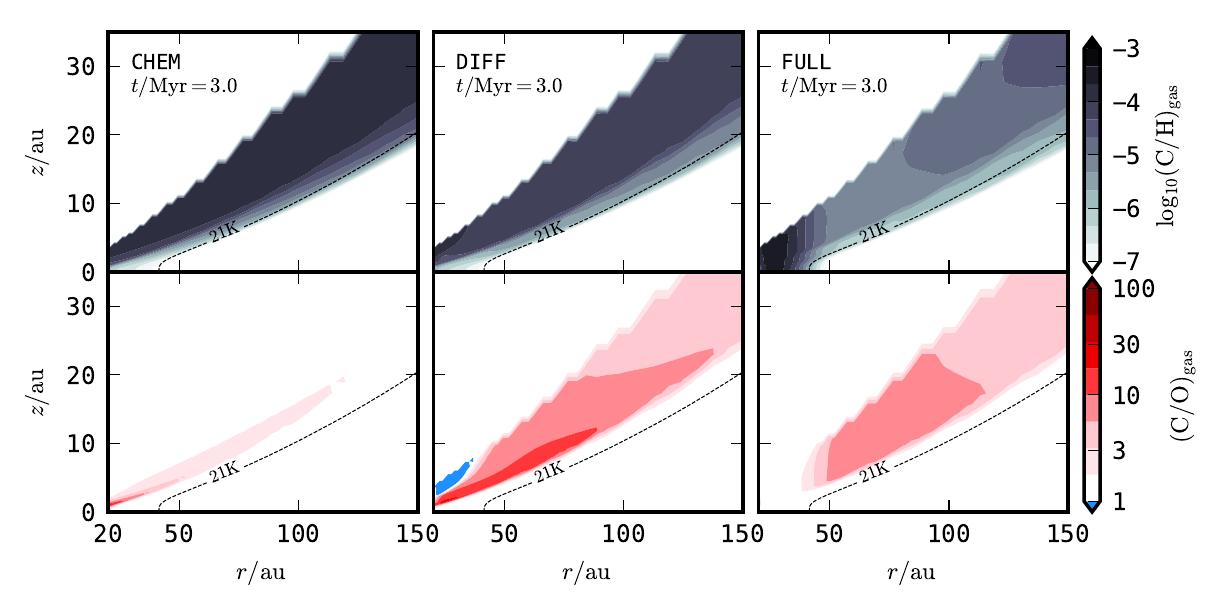}
\caption{Gas-phase C/H (top) and C/O (bottom) ratios after $3~\mathrm{~Myr}$ for the \texttt{CHEM} (left), \texttt{DIFF} (middle), and \texttt{FULL} (right) models.}
\label{fig:CoverO_g}
\end{figure*}

Figure~\ref{fig:CoverO_g} shows the gas-phase (C/H) and (C/O) ratios after $3\mathrm{~Myr}$ for \texttt{CHEM}, \texttt{DIFF}, and \texttt{FULL} simulations. The behavior seen for $\mathrm{(C/H)_{gas}}$ largely follows the \ce{CO} depletion pattern, as described for each simulation in Sect.~\ref{sec:results} in more detail. 

The story for $\mathrm{(C/O)_{gas}}$ is somewhat more involved. Initially, \ce{CO} dominates the carbon and oxygen reservoirs in the gas, resulting in $\mathrm{(C/O)_{gas}}\approx1$ at $t=0$. As a result of chemical processing, this ratio tends to increase with time for the pressures and temperatures we are considering. The reason is that of the chemical products considered here, \ce{CH4} is the most volatile (after \ce{CO}, see Table~\ref{tab:molecules}), its gas abundance pushing $\mathrm{(C/O)_{gas}}>1$ as the more oxygen-rich products (e.g., \ce{H2O}, \ce{CO2}) readily freeze out. This effect is similar to the one described in \citet[][Sect.~4.3.3]{schwarz2019}, who reported $\mathrm{(C/O)_{gas}}$ as high as ${\sim}10^8$ in the midplane at $r=12\mathrm{~au}$ after $6\mathrm{~Myr}$, when virtually all \ce{CO} had been processed.

In our simulations, the magnitude of the increase in $\mathrm{(C/O)_{gas}}$ and the extent of the region where it is seen depend on the transport processes that are included. In the static \texttt{CHEM} model (left panels of Fig.~\ref{fig:CoverO_g}), only a relatively small region shows $\mathrm{(C/O)_{gas}}>1$, since a significant fraction of the initial \ce{CO} is still present after $3~\mathrm{Myr}$ (see Fig.~\ref{fig:NG01}).

For the \texttt{DIFF} model, the combination of more efficient \ce{CO} removal combined with the upward transport of \ce{CH4} formed closer to the midplane results in a large portion of the WML having $\mathrm{(C/O)_{gas}}\sim10{-}30$. There is a small region in the upper, inner part of the disk that has $\mathrm{(C/O)_{gas}}<1$, due to sublimation of \ce{CO2} that formed at lower temperatures and was subsequently transported there.

The \texttt{FULL} model also results in carbon-rich gas in the WML, but the increase is not as extreme as for the \texttt{DIFF} model without pebble formation and evolution. There are 2 reasons for this. First, in the inner disk, the enhancement in \ce{CO} vapor interior to the \ce{CO} snowline (see Sect.~\ref{sec:NG07b}) dominates all other effects, forcing $\mathrm{(C/O)_{gas}}\approx1$. Thus, if radial drift is taking place, we would not expect to see the extremely high (C/O) ratios in the inner disk midplane gas as found by \citet{schwarz2019}. Further out, say, in the WML between $r\sim50{-}100\mathrm{~au}$, the sequestration of a substantial amount of \ce{CH4} ice on settled pebbles likely plays a role in the $\mathrm{(C/O)_{gas}}$ ratio being slightly lower than in the \texttt{DIFF} model, but still significantly ${\sim}3{-}10$.

Designed to be focused on the processing of \ce{CO}, the reduced chemical network we have used here does not consider further processing of \ce{CH4} and the formation of other, generally less volatile, hydrocarbons. As such, the abundance of \ce{CH4} at the end of our calculations, and the increase of $\mathrm{(C/O)_{gas}}$ associated with its appearance, should be treated as an upper limit. 

\subsubsection{Total volatile C/H and C/O}
In models without dust coagulation and pebble dynamics, the total volatile (i.e., gas + ice) C/O ratio does not change with time, since there is either no transport between neighbouring cells (for the static \texttt{CHEM} model) or because the dynamics of gas-phase species and small ice-carrying grains are essentially identical (for the \texttt{DIFF} calculation). For these models, the initial concentrations described in Table~\ref{tab:molecules} result in $(\mathrm{C/O})_\mathrm{gas+ice}\approx0.5$ and $(\mathrm{C/H})_\mathrm{gas+ice}\approx10^{-4}$ everywhere in the disk. 

\begin{figure*}
\centering
\includegraphics[clip=,width=1.\linewidth]{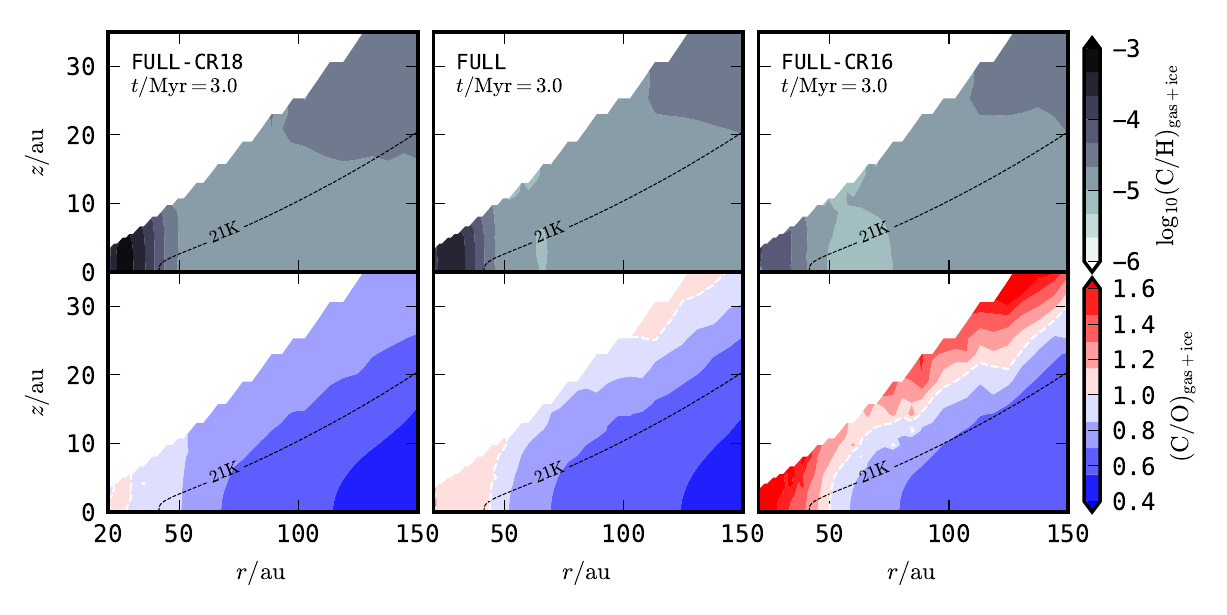}
\caption{Total (gas+ice) C/H (top) and C/O (bottom) ratios after $3~\mathrm{~Myr}$ for the \texttt{FULL-CR18} (left), \texttt{FULL} (middle), and \texttt{FULL-CR16} (right) models. The white dashed line shows $(\mathrm{C/O})_\mathrm{gas+ice}=1$. Note the different scales for the colorbar compared to Fig.~\ref{fig:CoverO_g}.}
\label{fig:CoverO_t}
\end{figure*}

When dust evolution is included, however, the dynamics of carbon and oxygen can differ depending on how they are partitioned between gas and ice, leading to spatial and temporal variations in the total elemental ratios \citep[e.g.,][]{oberg2016,booth2019}. Figure~\ref{fig:CoverO_t} shows the total\footnote{Excluding contributions from pebbles, since these are found only very close to the midplane.} volatile C/H and C/O ratios for the \texttt{FULL-CR18}, \texttt{FULL}, and \texttt{FULL-CR16} models (described earlier in Sect.~\ref{sec:substructures}).

The behavior shown by the $(\mathrm{C/H})_\mathrm{gas+ice}$ can be understood in terms of dust coagulation. As dust grows, especially in the regions exterior to the \ce{CO} snowline, a significant amount of carbon gets locked up in pebbles, reducing $(\mathrm{C/H})_\mathrm{gas+ice}$ on a timescale comparable to pebble formation. This process is slightly less effective in the disk surface layers, where the carbon in \ce{CO} and \ce{CH4} is in the gas-phase. Inside the \ce{CO} snowline, sublimation of inward drifting pebbles (e.g., Sect.~\ref{sec:NG07b}) leads to a $(\mathrm{C/H})_\mathrm{gas+ice}$ ratio that is elevated by a factor ${\sim}2{-}3$ compared to the initial value. For the \texttt{FULL-CR16} model, the values for $(\mathrm{C/H})_\mathrm{gas+ice}$ in the inner regions are lower, as the more rapid chemical processing makes it easier for pebbles to carry away carbon in the form of \ce{CO2}, and later \ce{CH4}, ice.

The variations in $(\mathrm{C/O})_\mathrm{gas+ice}$ are caused by a mix of chemical processing and dust evolution, and can be understood as follows. In the cold outer disk midplane, virtually all carbon and oxygen are frozen out and $(\mathrm{C/O})_\mathrm{gas+ice} \approx (\mathrm{C/O})_\mathrm{ice}$. Even as pebbles form and leave this region, the remaining ice dominates the carbon and oxygen budget, leaving $(\mathrm{C/O})_\mathrm{gas+ice}$ relatively unaffected. In the surface layers, and inside and around the \ce{CO} snowline, the picture is different. Here, initially, a significant amount of C and O are found in the gas, and the ice is generally more oxygen-rich than the gas, with the more volatile species (\ce{CO} and \ce{CH4}) having $\mathrm{C/O}>1$. As the oxygen-rich ice is removed over time, $(\mathrm{C/O})_\mathrm{gas+ice}$ slowly increases, and even becomes ${>}1$ as the abundance of \ce{CH4} becomes comparable to that of \ce{CO}. This happens faster in models with a higher $\zeta_\mathrm{CR}$, explaining the increasing $(\mathrm{C/O})_\mathrm{gas+ice}$ values going from \texttt{FULL-CR18}, to \texttt{FULL}, and \texttt{FULL-CR16}.

Elevated values $(\mathrm{C/O})_\mathrm{gas+ice}$ have been inferred from observation of multiple disks, with $(\mathrm{C/O})_\mathrm{gas+ice}$ ranging between 0.8 and 2.0 \citep{bergin2016, cleeves2018, miotello2019, legal2019, bergner2019}. This is comparable to the range of values found in the \texttt{FULL-CR18}, \texttt{FULL}, and \texttt{FULL-CR16} models. The models however, predict a strong correlation between CO abundance and $(\mathrm{C/O})_\mathrm{gas+ice}$ ratio in the surface layers, which so far has not been observed \citep{miotello2019}. Again, the lack of further processing of \ce{CH4} could lead us to over predicting the $(\mathrm{C/O})_\mathrm{gas+ice}$ in the top layers of the disk. Even so the dynamical and chemical processes studied here are at least in part responsible for the elevated $(\mathrm{C/O})_\mathrm{gas+ice}$ inferred from observations.

\subsection{Implications for giant planet compositions}\label{sec:giant_planets}
Following the seminal work of \citet{oberg2011}, many studies have sought to connect chemical abundances in the gas and solids, as found in the disk midplane, to the compositions of giant planet atmospheres \citep[e.g.,][]{oberg2016,booth2017,booth2019,cridland2019}. Recently, motivated by the theoretical and observational evidence that massive planets accrete gas primarily from the disk surface layers \citep{morbidelli2014,szulagyi2014,teague2019}, \citet{cridland2020} explored how gas accretion from reservoirs further up in the disk can alter the giant planet atmospheric composition, generally finding that the final C/O ratio was lower as the result of the accretion of oxygen-rich ices from between 1{-}3 scale-heights. The vertical profiles of the molecular abundances at the basis of the accretion model of \citet{cridland2020} were based on otherwise static disk model. It would then be interesting to explore how the changes predicted here (e.g., Fig.~\ref{fig:CoverO_t}) would impact the make-up of giant planets forming in the outer disk. However, only a relatively small fraction of the planet's final mass was accreted from the disk surface layers, especially for planets that (start) forming outside ${\sim}30\mathrm{~au}$ \citep[][Fig.~10]{cridland2020}. A more significant change in giant planet C/O ratios could come from the accretion of ice-rich pebbles in the disk midplane \citep[e.g.,][]{johansen2017}. Understanding the balance between radial drift and chemical processing timescales (e.g., Sect.~\ref{sec:substructures}) is then key, as pebbles that carry processed carbon in the form of \ce{CH3OH} or \ce{CO2} can hold on to their carbon for much longer than pebbles that are covered in more volatile \ce{CO} ice.

\subsection{Connection to CC-KBOs, (exo)comets, and interstellar interlopers}\label{sec:KBOs}
A more direct comparison can be made between the ice mantles in our simulations and the compositions of left-over planetesimals in the solar system and beyond. For example, \citet{eistrup2019} recently compared constraints on cometary compositions to the time and location dependent ice abundances in their static chemical model, finding that the majority of cometary properties can be explained by a formation location in the vicinity of the \ce{CO} ice-line.

The most direct connection is arguably possible for small bodies in the Cold Classical Kuiper Belt (CC-KB). This population of planetesimals, orbiting the sun between ${\sim}42$ and $47\mathrm{~au}$ on low-inclination, low-eccentricity orbits, is unique in that it contains some of the most pristine bodies in the solar system \citep{prialnik2020}. Moreover, the orbits of these bodies mostly undisturbed by the Neptune's outward migration \citep{nesvorny2015}, suggesting their current location very close to their formation location. Finally, there is compelling evidence that these bodies formed through the gentle collapse of cloud of pebble-size particles concentrated via the streaming instability \citep[e.g.,][]{nesvorny2019,mckinnon2020}, implying that their bulk composition at the time of formation resembles closely that of the pebble precursors. 

The best constraints on CC-KB object compositions come form the NASA New Horizons fly-by Arrokoth\footnote{Formerly known as (486958) 2014 MU$_{69}$.} early in 2019 \citep{stern2019}. While a body as small as Arrokoth cannot hold on to hyper-volatiles like \ce{CO} \citep{brown2012}, a surprising finding was the clear detection of \ce{CH3OH} ice, coupled with a lack of strong \ce{H2O} ice features in the reflected NIR spectrum \citep{grundy2020}. While it is unclear to what extent these surface features represent the bulk composition \citep[see discussion in][]{grundy2020}, we can attempt to compare these findings to results in our models.

\begin{figure*}
\centering
\includegraphics[clip=,width=1.\linewidth]{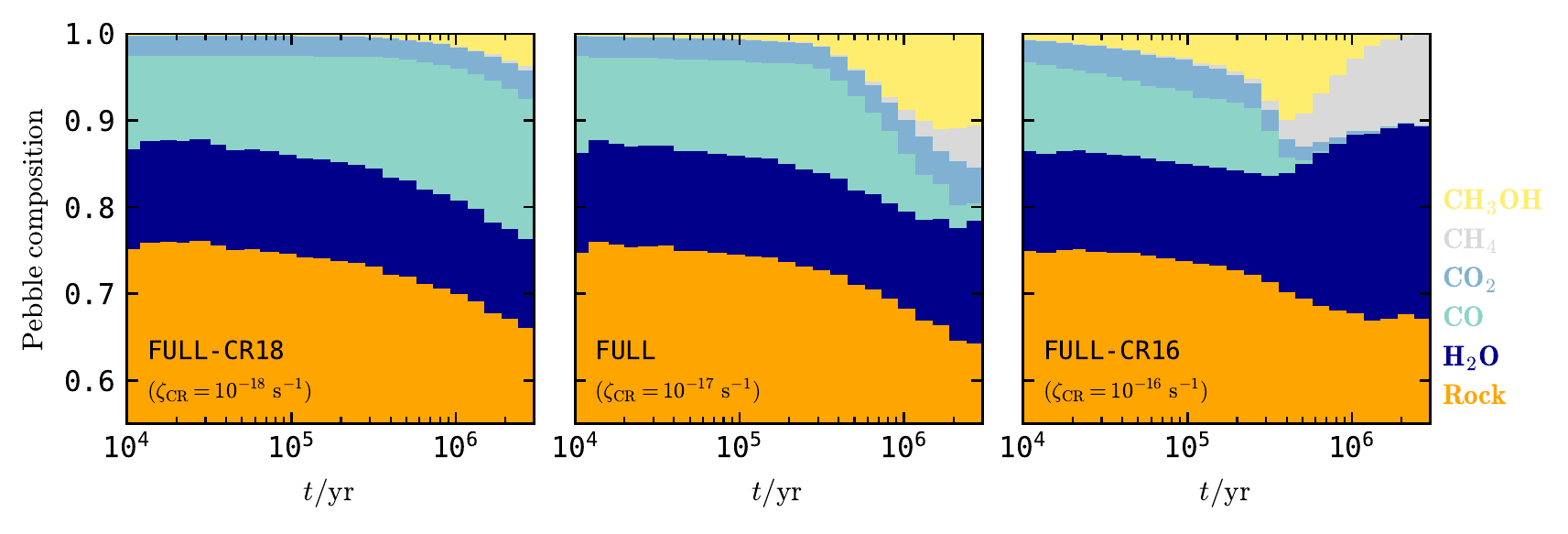}
\caption{Normalized mass-weighted compositions of pebbles located between $r=42$ and $47\mathrm{~au}$ as a function of time, for the \texttt{FULL-CR18}, \texttt{FULL}, and \texttt{FULL-CR16} models.}
\label{fig:pebbles45}
\end{figure*}

Figure~\ref{fig:pebbles45} shows the (mass-weighted) composition of all pebble particles present between $42{-}47\mathrm{~au}$ (i.e., the location of the CC-KB) as a function of time for three models with different cosmic ray ionization rates. For the model with $\zeta_\mathrm{CR}=10^{-18}\mathrm{~s^{-1}}$, grain surface chemistry is slow and the pebbles' ice mantles are always composed of a large amount of \ce{CO} ice, with $\lesssim5\%$ \ce{CH3OH} ice. For the standard \texttt{FULL} model (middle panel), the highest concentration of methanol ice (relative to water ice) is reached after ${\sim}2\mathrm{~Myr}$. In the model with a high cosmic ray rate (right panel), a similar situation is reached already after ${\sim}0.4\mathrm{~Myr}$, after which the contribution of \ce{CH3OH} begins to diminish. 

The current mass of the entire CC-KB population is estimated to be ${\sim}3\times10^{-4}M_\oplus$ \citep{fraser2014}, and region probably only suffered a factor of ${\sim}2$ dynamical depletion during Neptune's outward migration \citep{nesvorny2015}. Thus, even after $3\mathrm{~Myr}$ in our simulations the region between $42{-}47\mathrm{~au}$ contains enough mass in pebbles to produce a CC-KB like population of planetesimals (see Fig.~\ref{fig:pebble_masses}). 

Looking beyond the solar system, observations of second-generation \ce{CO} gas (and its derivatives) in old (${>}10{-}100\mathrm{~Myr}$) and otherwise gas-poor debris disks are being employed to constrain the combined \ce{CO}+\ce{CO2} mass fraction of planetesimals at the top of the collisional cascade \citep[][and references therein]{wyatt2020}. For example, \citet{matra2017} constrain the combined \ce{CO}+\ce{CO2} ice mass fraction to lie between ${\sim}0.3{-}30\%$ for the planetesimals in HD181327 and Fomalhaut. As these constraints become tighter and more systems are added, comparing such constraints to models like the ones presented here, and looking for trends with e.g., stellar luminosity or planetesimal belt location will undoubtedly provide valuable insights.

Lastly, interstellar interloper 2I/Borisov appears to be very rich in \ce{CO} ice compared to solar system comets \citep{bodewits2020,cordiner2020}, which has been taken as evidence for its formation exterior to the \ce{CO} snowline. In the context of the models presented here, we would argue that, if indeed \ce{CO}-rich, 2I/Borisov and the exo-comets in e.g. Fomalhaut not only formed exterior to the \ce{CO} snowline in their systems, but also early during disk evolution, before significant processing of \ce{CO} ice on grain surfaces took place. 

We end by stressing that, in the simulations presented here, the local dust-to-gas ratio in the midplane is not readily enhanced sufficiently to result in planetesimal formation via the streaming instability \citep[see also][]{krijt2016a,krijt2018}. It appears then that additional physics such as disk photoevaporation \citep{carrera2017} or the presence of (short-lived) pressure traps \citep{lenz2019} are needed to facilitate planetesimal formation outside of $r\gtrsim10{-}100\mathrm{~au}$.

\section{Summary}
Motivated by recent observational constraints on the disappearance of gas-phase \ce{CO} from the warm molecular layers of gas-rich protoplanetary disks \citep{bergin2018,zhang2019,zhang2020a,zhang2020b}, we have self-consistently modeled the two sets of processes generally though to be responsible: chemical processing \citep{bosman2018b,schwarz2018} and sequestration in the midplane \citep{kama2016,krijt2018}.

Our main findings are as follows:
\begin{enumerate}

    \item In our fiducial model with $\zeta_\mathrm{CR}=10^{-17}\mathrm{~s^{-1}}$, chemical processing of \ce{CO} combined with ice sequestration in the disk midplane, result in \ce{CO} depletion factors of ${\sim}100$ in the outer disk warm molecular layer after $t=3\mathrm{~Myr}$ (Fig.~\ref{fig:NG07b}).
    
    \item Models that do not include chemical processing, dust coagulation and dynamics, or employ a lower cosmic ray ionization rate, do not reach similar depletion factors in the outer disk (Fig.~\ref{fig:profiles} and \ref{fig:profiles_2}).
    
    \item Models that include pebble drift result in an enhanced \ce{CO} abundance interior to the midplane \ce{CO} snowline (Figs.~\ref{fig:NG06b}, \ref{fig:NG07b}, and \ref{fig:profiles}), unless chemical processing operates on timescales shorter than or comparable to disk-wide radial drift (Fig.~\ref{fig:profiles_2}).
    
    \item Including chemical processing and dust dynamics results in elevated $\mathrm{(C/O)_\mathrm{gas}}\sim3{-}10$ in the warm molecular layer in the outer disk, while inward pebble drift and \ce{CO} sublimation force $\mathrm{(C/O)_\mathrm{gas}}\approx1$ closer to the star (Fig.~\ref{fig:CoverO_g}).
    
    \item Dust evolution and material transport also alter the total volatile elemental abundances ratio, typically leading to $\mathrm{(C/O)_\mathrm{gas+ice}}\gtrsim1$ inside the midplane CO snowline and parts of the warm molecular layer at larger radii (Fig.~\ref{fig:CoverO_t}).
    
    \item The removal of solids and warm \ce{CO} from the outer disk on ${\sim}\mathrm{Myr}$ timescales complicates the use of their emission as a bulk disk mass tracer (Fig.~\ref{fig:disk_mass}).
    
    \item Pebble compositions predicted by models like the ones presented here can be compared to knowledge of planetesimal compositions in the Kuiper belt and (extra-solar) comets to constrain the timing and location of their formation (Sect.~\ref{sec:KBOs}), although planetesimal formation is not yet treated self-consistently in this work.
    
\end{enumerate}
More broadly, our results highlight the need for studying the interplay between physical and chemical processes in gas-rich protoplanetary disks, and demonstrate how molecular emission originating from the disk warm molecular layer can be used to constrain the growth and drift of solids in the disk midplane.

\acknowledgments
SK would like to thank Mihkel Kama, Ilaria Pascucci, Andrea Banzatti, Dmitry Semenov, and D\'{a}niel Apai for stimulating and encouraging discussions. SK, KZ, and KS acknowledge the support of NASA through Hubble Fellowship grants HST-HF2-51394.001, HST-HF2-51401.001, and HST-HF2-51419.001, respectively, awarded by the Space Telescope Science Institute, which is operated by the Association of Universities for Research in Astronomy, Inc., for NASA, under contract NAS5-26555. ADB acknowledges support from NSF Grant\#1907653. The results reported herein benefited from collaborations and/or information exchange within NASA's Nexus for Exoplanet System Science (NExSS) research coordination network sponsored by NASA’s Science Mission Directorate. 

\noindent\emph{Software:} \texttt{matplotlib} \citep{hunter2007}.

\newpage

\appendix
\section{A. Reduced chemical network}\label{sec:appendix_network}

In \citet{bosman2018b} three pathways for the conversion of \ce{CO} into other species have been studied using a full chemical network. As combining a full chemical network within a dynamical simulation is prohibitively expensive, we have created a reduced network that traces the major carbon carriers (\ce{CO}, \ce{CO2} and \ce{CH3OH}) as well as water and exhibits the same behaviour as the full chemical network in these species. Table~\ref{tab:reac_all} shows the reactions included in the reduced network. All the rates are calculated as in \citep{bosman2018b}, and coefficients are taken from \cite{McElroy2013}, \citet{Heays2017} and \citet{Garrod2008}. All chemistry is ultimately driven by cosmic-rays, as such the cosmic-ray ionisation rate is critical parameter. More directly, \ce{CO} conversion is dominated by three reactions\footnote{In the Appendix only, (s) is used to denote molecules/species frozen out on grain surfaces.}:
\begin{subequations}
\begin{equation}
\label{eq:COHe}
R_\mathrm{\ce{CO + He^+}}:\ce{CO + He^+ -> C^+ + O}
\end{equation}
\begin{equation}
\label{eq:COH}
R_\mathrm{\ce{CO + H}}:\ce{CO(s) + H(s)+ -> HCO(s)}
\end{equation}
\begin{equation}
\label{eq:COOH}
R_\mathrm{\ce{CO + OH}}:\ce{CO(s) + OH(s)+ -> CO2(s) + H(s)}
\end{equation}
\end{subequations}
For reaction~\ref{eq:COHe}, it is assumed that all carbon ends up as \ce{CH4}, while for reaction~\ref{eq:COH}, \ce{HCO} is assumed quickly get further hydrogenated to \ce{CH3OH}. 
The abundances of \ce{He^{+}}, \ce{H(s)} and \ce{OH(s)} are thus necessary for the calculation of the evolution of the \ce{CO} abundance. For each of these three species, we make the simplification that they are in equilibrium and that their abundance is thus given by equating the formation and destruction reactions for these species. 

\ce{He^{+}} is only created by direct cosmic-ray ionisation of \ce{He} and thus has a constant formation rate. \ce{He^{+}} is destroyed by reactions with gaseous \ce{CO}, \ce{CH4} and \ce{N2}. The abundance for the first two species is tracked, while \ce{N2} is assumed to be constant at an abundance of $ 10^{-5}$ w.r.t \ce{H}. 

The \ce{H(s)} abundance is a bit more difficult to calculate. To get the abundance of \ce{H} on the ice, the abundance of \ce{H} in the gas-phase needs to be calculated first. In the gas-phase \ce{H} is created by cosmic-ray ionisation of \ce{H2} and subsequent reactions. Depending on the dominant pathway, between 2 and 4 \ce{H} atoms are created after one \ce{H2} ionisation. For the reduced network we use an average value of 2.4 \ce{H} per \ce{H2} ionisation which fits well with a broad range of conditions that are considered here. This reaction dominates the production Atomic \ce{H} in the gas-phase. Atomic \ce{H} get removed from the gas-phase, by the production of \ce{H2}, for which we follow \citep{Cazaux2004}, and the freeze-out of \ce{H} onto the dust grain. Balancing these rates gives an abundance for \ce{H} in the gas-phase and thus a freeze-out rate of \ce{H}. Sublimation of \ce{H} is not included as the full chemical network predicts that more than 99\% of the H that freezes out on the grains will react on the grain. 

On the grain, atomic \ce{H} can react with a number of species. Besides the reactions of \ce{H(s)} with \ce{O}, \ce{OH} and \ce{CO}, catalytic formation of \ce{H2} on the ice is also considered using reaction of the form: \ce{XH(s) + H(s) -> X(s) + H2(s)} and 
\ce{X(s) + H(s) -> XH(s)}, where the former is the rate limiting step. In the full chemical network reactions of \ce{H} with \ce{H2S} was the dominant pathway in most of parameter space. It is assumed that the \ce{H2S(s)} abundance is $2\times 10^{-8}$. To calculate the \ce{H2} formation rate, three times the rate of \ce{H2S(s) + H(s)} is assumed, to correct for the \ce{SH(s) + H(s)} reaction as well as catalytic reactions with other species on the ice. 

Finally the \ce{OH(s)} abundance needs to be calculated. It is assumed that all \ce{O} that is produced from the dissociation of \ce{CO}, \ce{CO2} and \ce{H2O}, is quickly turned into \ce{OH(s)}, further \ce{OH(s)} is produced in the dissociation of \ce{CH3OH(s)}. \ce{OH(s)} can then react with \ce{H(s)} and \ce{H2(s)}, forming \ce{H2O} and \ce{CO(s)}, forming \ce{CO2(s)}. 

\begin{figure}
    \centering
    \includegraphics[width = \hsize]{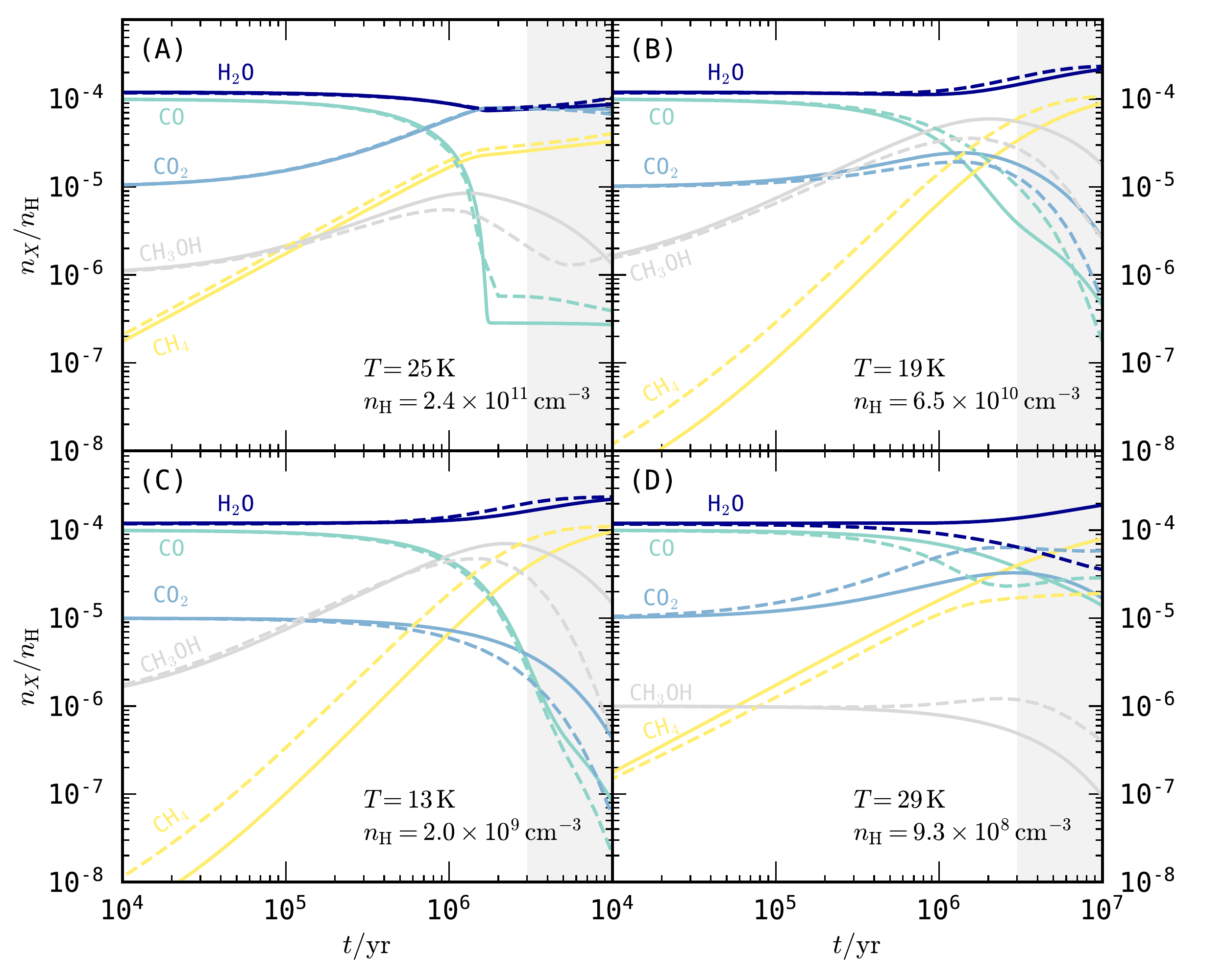}
    \caption{Comparison between the approximate model (solid) and the full chemical model (dashed), for four different combinations of gas temperature and density, corresponding (roughly) to locations A{-}D shown in Fig.~\ref{fig:NG01}. All curves show combined gas + ice abundances, for $\zeta_\mathrm{CR}=10^{-17}\mathrm{~s^{-1}}$, a constant dust-to-gas ratio of $10^{-2}$, and no grain growth. The grey area indicates times $t > 3\mathrm{~Myr}$. The approximate model closely follows the \ce{CO} abundance as well as the abundance of the major carbon carrier, all within a factor of 2.}
    \label{fig:full_comparison}
\end{figure}

Calculating the abundances of these three key intermediate species allows for the calculation of the evolution of the four traced species, \ce{CO}, \ce{CO2}, \ce{H2O} and \ce{CH3OH}. As the evolution of these species is slow, and the abundance of the intermediate species only depends on the traced species, large times steps (in the order of kyr) can be taken in the computation of the chemical evolution. As the chemical times-scales speed up when the \ce{CO} abundance drops, we stop the conversion of \ce{CO} when its abundance is below $10^{-8}$. Fig.~\ref{fig:full_comparison} shows the comparison between the full chemical model and the approximate model for four different physical conditions. These are the same conditions as studied in \citet{bosman2018b} and sample all CO conversion pathways. 

\begin{table}
\centering
\caption{\label{tab:reac_all}Reactions included}
\begin{tabular}{rl l}
\hline
\hline 
\multicolumn{2}{c}{Reaction} & Comments \\
\hline 
\ce{H + H }&\ce{-> H2} & \ce{H2} grain surface formation according to \citet{Cazaux2004}  \\
\ce{HX(s) + H(s)}&\ce{-> X(s) + H2(s)} & \multirow{2}{*}{catalytic \ce{H2} formation on the grains \citep{tielens1982}}\\
\ce{X(s) + H(s)}&\ce{-> HX(s)} &  \\ 
\ce{H2O(s) + $\gamma$}&\ce{-> O(s) + H(s) + H(s)} & Cosmic ray induced photo dissociation \citep{Heays2017} \\
\ce{CO2(s) + $\gamma$}&\ce{-> CO(s) + O(s)} & Cosmic ray induced photo dissociation \citep{Heays2017} \\
\ce{CH3OH(s) + $\gamma$}&\ce{-> CH3(s) + OH(s)} & Cosmic ray induced photo dissociation \citep{Heays2017} \\
\ce{nH2 + 2nCR}&\ce{-> 2nH} & \ce{H2} dissociation due to cosmic rays efficiency n is between 2 and 4 \\
\ce{He + CR}&\ce{-> He^+ + e^-} & Cosmic ray ionisation of \ce{He} \\
\ce{He^{+} + CO}&\ce{-> C^+ + O + He }& \\
\ce{He^{+} + N2}&\ce{-> N2^+ + He }& \multirow{3}{5cm}{Reactions of \ce{He^+}}\\
\ce{He^{+} + CH4}&\ce{-> CH4^+ He}& \\
\ce{O(s) + H(s)}&\ce{-> OH(s)}  &Hydrogenation of oxygen to \ce{OH}, assumed to be instantaneous \\
\ce{CO(s) + 4 H(s)}&\ce{-> CH3OH(s)} & Hydrogenation of \ce{CO}, the initial step is assumed to be rate limiting \\
\ce{CO(s) + OH(s)}&\ce{-> CO2(s)} & \ce{CO2} formation on the grain \\
\ce{OH(s) + H(s)}&\ce{-> H2O(s)} & \ce{H2O} formation on the grain \\
\ce{OH(s) + H2(s)}&\ce{-> H2O(s) + H(s)} & \ce{H2O} formation on the grain \\
\ce{C^+ + xH2 + e^{-}}&\ce{-> CHx + xH}  & \multirow{3}{5cm}{Formation of \ce{CH4} through ion-molecule and hydrogenation reactions}\\
\ce{CHx}&\ce{->  CHx(s)} & \\ 
\ce{CHx(s) + (4-x)H}&\ce{->  CH4(s)} & \\
\hline
\end{tabular}
\end{table}

\bibliographystyle{aa}
\bibliography{refs}

\begin{thebibliography}{112}
\expandafter\ifx\csname natexlab\endcsname\relax\def\natexlab#1{#1}\fi

\bibitem[{{Aikawa} {et~al.}(1996){Aikawa}, {Miyama}, {Nakano}, \&
  {Umebayashi}}]{aikawa1996}
{Aikawa}, Y., {Miyama}, S.~M., {Nakano}, T., \& {Umebayashi}, T. 1996, \apj,
  467, 684

\bibitem[{{Aikawa} {et~al.}(2002){Aikawa}, {van Zadelhoff}, {van Dishoeck}, \&
  {Herbst}}]{aikawa2002}
{Aikawa}, Y., {van Zadelhoff}, G.~J., {van Dishoeck}, E.~F., \& {Herbst}, E.
  2002, \aap, 386, 622

\bibitem[{{Andrews}(2020)}]{andrews2020}
{Andrews}, S.~M. 2020, arXiv e-prints, arXiv:2001.05007

\bibitem[{{Andrews} {et~al.}(2012){Andrews}, {Wilner}, {Hughes}, {Qi},
  {Rosenfeld}, {{\"O}berg}, {Birnstiel}, {Espaillat}, {Cieza}, {Williams},
  {Lin}, \& {Ho}}]{andrews2012}
{Andrews}, S.~M., {Wilner}, D.~J., {Hughes}, A.~M., {et~al.} 2012, \apj, 744,
  162

\bibitem[{{Ansdell} {et~al.}(2018){Ansdell}, {Williams}, {Trapman}, {van
  Terwisga}, {Facchini}, {Manara}, {van der Marel}, {Miotello}, {Tazzari},
  {Hogerheijde}, {Guidi}, {Testi}, \& {van Dishoeck}}]{ansdell2018}
{Ansdell}, M., {Williams}, J.~P., {Trapman}, L., {et~al.} 2018, \apj, 859, 21

\bibitem[{{Ansdell} {et~al.}(2016){Ansdell}, {Williams}, {van der Marel},
  {Carpenter}, {Guidi}, {Hogerheijde}, {Mathews}, {Manara}, {Miotello},
  {Natta}, {Oliveira}, {Tazzari}, {Testi}, {van Dishoeck}, \& {van
  Terwisga}}]{ansdell2016}
{Ansdell}, M., {Williams}, J.~P., {van der Marel}, N., {et~al.} 2016, \apj,
  828, 46

\bibitem[{{Bae} {et~al.}(2019){Bae}, {Zhu}, {Baruteau}, {Benisty}, {Dullemond},
  {Facchini}, {Isella}, {Keppler}, {P{\'e}rez}, \& {Teague}}]{bae2019}
{Bae}, J., {Zhu}, Z., {Baruteau}, C., {et~al.} 2019, \apjl, 884, L41

\bibitem[{{Bergin} {et~al.}(2014){Bergin}, {Cleeves}, {Crockett}, \&
  {Blake}}]{bergin2014}
{Bergin}, E.~A., {Cleeves}, L.~I., {Crockett}, N., \& {Blake}, G.~A. 2014,
  Faraday Discussions, 168, 61

\bibitem[{{Bergin} {et~al.}(2013){Bergin}, {Cleeves}, {Gorti}, {Zhang},
  {Blake}, {Green}, {Andrews}, {Evans}, {Henning}, {{\"O}berg}, {Pontoppidan},
  {Qi}, {Salyk}, \& {van Dishoeck}}]{bergin2013}
{Bergin}, E.~A., {Cleeves}, L.~I., {Gorti}, U., {et~al.} 2013, \nat, 493, 644

\bibitem[{{Bergin} {et~al.}(2016){Bergin}, {Du}, {Cleeves}, {Blake}, {Schwarz},
  {Visser}, \& {Zhang}}]{bergin2016}
{Bergin}, E.~A., {Du}, F., {Cleeves}, L.~I., {et~al.} 2016, \apj, 831, 101

\bibitem[{{Bergin} \& {Williams}(2017)}]{bergin2018}
{Bergin}, E.~A. \& {Williams}, J.~P. 2017, Astrophysics and Space Science
  Library, Vol. 445, {The Determination of Protoplanetary Disk Masses}, ed.
  M.~{Pessah} \& O.~{Gressel}, 1

\bibitem[{{Bergner} {et~al.}(2020){Bergner}, {Oberg}, {Bergin}, {Andrews},
  {Blake}, {Carpenter}, {Cleeves}, {Guzman}, {Huang}, {Jorgensen}, {Qi},
  {Schwarz}, {Williams}, \& {Wilner}}]{bergner2020}
{Bergner}, J.~B., {Oberg}, K.~I., {Bergin}, E.~A., {et~al.} 2020, arXiv
  e-prints, arXiv:2006.12584

\bibitem[{{Bergner} {et~al.}(2019){Bergner}, {{\"O}berg}, {Bergin}, {Loomis},
  {Pegues}, \& {Qi}}]{bergner2019}
{Bergner}, J.~B., {{\"O}berg}, K.~I., {Bergin}, E.~A., {et~al.} 2019, \apj,
  876, 25

\bibitem[{{Birnstiel} {et~al.}(2012){Birnstiel}, {Klahr}, \&
  {Ercolano}}]{birnstiel2012}
{Birnstiel}, T., {Klahr}, H., \& {Ercolano}, B. 2012, \aap, 539, A148

\bibitem[{{Bodewits} {et~al.}(2020){Bodewits}, {Noonan}, {Feldman},
  {Bannister}, {Farnocchia}, {Harris}, {Li}, {Mandt}, {Parker}, \&
  {Xing}}]{bodewits2020}
{Bodewits}, D., {Noonan}, J.~W., {Feldman}, P.~D., {et~al.} 2020, Nature
  Astronomy

\bibitem[{{Booth} {et~al.}(2017){Booth}, {Clarke}, {Madhusudhan}, \&
  {Ilee}}]{booth2017}
{Booth}, R.~A., {Clarke}, C.~J., {Madhusudhan}, N., \& {Ilee}, J.~D. 2017,
  \mnras, 469, 3994

\bibitem[{{Booth} \& {Ilee}(2019)}]{booth2019}
{Booth}, R.~A. \& {Ilee}, J.~D. 2019, \mnras, 487, 3998

\bibitem[{{Bosman} \& {Banzatti}(2019)}]{bosman2019}
{Bosman}, A.~D. \& {Banzatti}, A. 2019, \aap, 632, L10

\bibitem[{{Bosman} {et~al.}(2018{\natexlab{a}}){Bosman}, {Tielens}, \& {van
  Dishoeck}}]{bosman2018a}
{Bosman}, A.~D., {Tielens}, A. G.~G.~M., \& {van Dishoeck}, E.~F.
  2018{\natexlab{a}}, \aap, 611, A80

\bibitem[{{Bosman} {et~al.}(2018{\natexlab{b}}){Bosman}, {Walsh}, \& {van
  Dishoeck}}]{bosman2018b}
{Bosman}, A.~D., {Walsh}, C., \& {van Dishoeck}, E.~F. 2018{\natexlab{b}},
  \aap, 618, A182

\bibitem[{{Boyden} \& {Eisner}(2020)}]{boyden2020}
{Boyden}, R.~D. \& {Eisner}, J.~A. 2020, \apj, 894, 74

\bibitem[{{Brown}(2012)}]{brown2012}
{Brown}, M.~E. 2012, Annual Review of Earth and Planetary Sciences, 40, 467

\bibitem[{{Carrera} {et~al.}(2017){Carrera}, {Gorti}, {Johansen}, \&
  {Davies}}]{carrera2017}
{Carrera}, D., {Gorti}, U., {Johansen}, A., \& {Davies}, M.~B. 2017, \apj, 839,
  16

\bibitem[{{Cazaux} \& {Tielens}(2004)}]{Cazaux2004}
{Cazaux}, S. \& {Tielens}, A.~G.~G.~M. 2004, \apj, 604, 222

\bibitem[{{Ciesla}(2009)}]{ciesla2009}
{Ciesla}, F.~J. 2009, \icarus, 200, 655

\bibitem[{{Ciesla} \& {Cuzzi}(2006)}]{ciesla2006}
{Ciesla}, F.~J. \& {Cuzzi}, J.~N. 2006, \icarus, 181, 178

\bibitem[{{Cleeves} {et~al.}(2013){Cleeves}, {Adams}, \&
  {Bergin}}]{cleeves2013}
{Cleeves}, L.~I., {Adams}, F.~C., \& {Bergin}, E.~A. 2013, \apj, 772, 5

\bibitem[{{Cleeves} {et~al.}(2015){Cleeves}, {Bergin}, {Qi}, {Adams}, \&
  {{\"O}berg}}]{cleeves2015}
{Cleeves}, L.~I., {Bergin}, E.~A., {Qi}, C., {Adams}, F.~C., \& {{\"O}berg},
  K.~I. 2015, \apj, 799, 204

\bibitem[{{Cleeves} {et~al.}(2018){Cleeves}, {{\"O}berg}, {Wilner}, {Huang},
  {Loomis}, {Andrews}, \& {Guzman}}]{cleeves2018}
{Cleeves}, L.~I., {{\"O}berg}, K.~I., {Wilner}, D.~J., {et~al.} 2018, \apj,
  865, 155

\bibitem[{{Cordiner} {et~al.}(2020){Cordiner}, {Milam}, {Biver},
  {Bockel{\'e}e-Morvan}, {Roth}, {Bergin}, {Jehin}, {Remijan}, {Charnley},
  {Mumma}, {Boissier}, {Crovisier}, {Paganini}, {Kuan}, \&
  {Lis}}]{cordiner2020}
{Cordiner}, M.~A., {Milam}, S.~N., {Biver}, N., {et~al.} 2020, Nature Astronomy

\bibitem[{{Cridland} {et~al.}(2020){Cridland}, {Bosman}, \& {van
  Dishoeck}}]{cridland2020}
{Cridland}, A.~J., {Bosman}, A.~D., \& {van Dishoeck}, E.~F. 2020, \aap, 635,
  A68

\bibitem[{{Cridland} {et~al.}(2019){Cridland}, {Eistrup}, \& {van
  Dishoeck}}]{cridland2019}
{Cridland}, A.~J., {Eistrup}, C., \& {van Dishoeck}, E.~F. 2019, \aap, 627,
  A127

\bibitem[{{Cuzzi} \& {Zahnle}(2004)}]{cuzzi2004}
{Cuzzi}, J.~N. \& {Zahnle}, K.~J. 2004, \apj, 614, 490

\bibitem[{{Dodson-Robinson} {et~al.}(2018){Dodson-Robinson}, {Evans}, {Ramos},
  {Yu}, \& {Willacy}}]{dodson-robinson2018}
{Dodson-Robinson}, S.~E., {Evans}, Neal~J., I., {Ramos}, A., {Yu}, M., \&
  {Willacy}, K. 2018, \apjl, 868, L37

\bibitem[{{Dullemond} {et~al.}(2020){Dullemond}, {Isella}, {Andrews},
  {Skobleva}, \& {Dzyurkevich}}]{dullemond2020}
{Dullemond}, C.~P., {Isella}, A., {Andrews}, S.~M., {Skobleva}, I., \&
  {Dzyurkevich}, N. 2020, \aap, 633, A137

\bibitem[{{Dutrey} {et~al.}(2017){Dutrey}, {Guilloteau}, {Pi{\'e}tu},
  {Chapillon}, {Wakelam}, {Di Folco}, {Stoecklin}, {Denis-Alpizar}, {Gorti},
  {Teague}, {Henning}, {Semenov}, \& {Grosso}}]{dutrey2017}
{Dutrey}, A., {Guilloteau}, S., {Pi{\'e}tu}, V., {et~al.} 2017, \aap, 607, A130

\bibitem[{{Eisner} {et~al.}(2016){Eisner}, {Bally}, {Ginsburg}, \&
  {Sheehan}}]{eisner2016}
{Eisner}, J.~A., {Bally}, J.~M., {Ginsburg}, A., \& {Sheehan}, P.~D. 2016,
  \apj, 826, 16

\bibitem[{{Eistrup} {et~al.}(2018){Eistrup}, {Walsh}, \& {van
  Dishoeck}}]{eistrup2018}
{Eistrup}, C., {Walsh}, C., \& {van Dishoeck}, E.~F. 2018, \aap, 613, A14

\bibitem[{{Eistrup} {et~al.}(2019){Eistrup}, {Walsh}, \& {van
  Dishoeck}}]{eistrup2019}
{Eistrup}, C., {Walsh}, C., \& {van Dishoeck}, E.~F. 2019, \aap, 629, A84

\bibitem[{{Estrada} {et~al.}(2016){Estrada}, {Cuzzi}, \&
  {Morgan}}]{estrada2016}
{Estrada}, P.~R., {Cuzzi}, J.~N., \& {Morgan}, D.~A. 2016, \apj, 818, 200

\bibitem[{{Favre} {et~al.}(2013){Favre}, {Cleeves}, {Bergin}, {Qi}, \&
  {Blake}}]{favre2013}
{Favre}, C., {Cleeves}, L.~I., {Bergin}, E.~A., {Qi}, C., \& {Blake}, G.~A.
  2013, \apjl, 776, L38

\bibitem[{{Flaherty} {et~al.}(2015){Flaherty}, {Hughes}, {Rosenfeld},
  {Andrews}, {Chiang}, {Simon}, {Kerzner}, \& {Wilner}}]{flaherty2015}
{Flaherty}, K.~M., {Hughes}, A.~M., {Rosenfeld}, K.~A., {et~al.} 2015, \apj,
  813, 99

\bibitem[{{Fraser} {et~al.}(2014){Fraser}, {Brown}, {Morbidelli}, {Parker}, \&
  {Batygin}}]{fraser2014}
{Fraser}, W.~C., {Brown}, M.~E., {Morbidelli}, A.~r., {Parker}, A., \&
  {Batygin}, K. 2014, \apj, 782, 100

\bibitem[{{Furuya} \& {Aikawa}(2014)}]{furuya2014}
{Furuya}, K. \& {Aikawa}, Y. 2014, \apj, 790, 97

\bibitem[{{Garrod} {et~al.}(2008){Garrod}, {Widicus Weaver}, \&
  {Herbst}}]{Garrod2008}
{Garrod}, R.~T., {Widicus Weaver}, S.~L., \& {Herbst}, E. 2008, \apj, 682, 283

\bibitem[{{Grundy} {et~al.}(2020){Grundy}, {Bird}, {Britt}, {Cook},
  {Cruikshank}, {Howett}, {Krijt}, {Linscott}, {Olkin}, {Parker}, {Protopapa},
  {Ruaud}, {Umurhan}, {Young}, {Dalle Ore}, {Kavelaars}, {Keane}, {Pendleton},
  {Porter}, {Scipioni}, {Spencer}, {Stern}, {Verbiscer}, {Weaver}, {Binzel},
  {Buie}, {Buratti}, {Cheng}, {Earle}, {Elliott}, {Gabasova}, {Gladstone},
  {Hill}, {Horanyi}, {Jennings}, {Lunsford}, {McComas}, {McKinnon}, {McNutt},
  {Moore}, {Parker}, {Quirico}, {Reuter}, {Schenk}, {Schmitt}, {Showalter},
  {Singer}, {Weigle}, \& {Zangari}}]{grundy2020}
{Grundy}, W.~M., {Bird}, M.~K., {Britt}, D.~T., {et~al.} 2020, Science, 367,
  aay3705

\bibitem[{{Heays} {et~al.}(2017){Heays}, {Bosman}, \& {van
  Dishoeck}}]{Heays2017}
{Heays}, A.~N., {Bosman}, A.~D., \& {van Dishoeck}, E.~F. 2017, \aap, 602, A105

\bibitem[{{Henning} \& {Semenov}(2013)}]{henning2013}
{Henning}, T. \& {Semenov}, D. 2013, Chemical Reviews, 113, 9016

\bibitem[{{Huang} {et~al.}(2018){Huang}, {Andrews}, {Dullemond}, {Isella},
  {P{\'e}rez}, {Guzm{\'a}n}, {{\"O}berg}, {Zhu}, {Zhang}, {Bai}, {Benisty},
  {Birnstiel}, {Carpenter}, {Hughes}, {Ricci}, {Weaver}, \&
  {Wilner}}]{huang2018}
{Huang}, J., {Andrews}, S.~M., {Dullemond}, C.~P., {et~al.} 2018, \apjl, 869,
  L42

\bibitem[{{Hunter}(2007)}]{hunter2007}
{Hunter}, J.~D. 2007, Computing in Science Engineering, 9, 90

\bibitem[{{Isella} {et~al.}(2018){Isella}, {Huang}, {Andrews}, {Dullemond},
  {Birnstiel}, {Zhang}, {Zhu}, {Guzm{\'a}n}, {P{\'e}rez}, {Bai}, {Benisty},
  {Carpenter}, {Ricci}, \& {Wilner}}]{isella2018}
{Isella}, A., {Huang}, J., {Andrews}, S.~M., {et~al.} 2018, \apjl, 869, L49

\bibitem[{{Johansen} \& {Lambrechts}(2017)}]{johansen2017}
{Johansen}, A. \& {Lambrechts}, M. 2017, Annual Review of Earth and Planetary
  Sciences, 45, 359

\bibitem[{{Kama} {et~al.}(2016){Kama}, {Bruderer}, {van Dishoeck},
  {Hogerheijde}, {Folsom}, {Miotello}, {Fedele}, {Belloche}, {G{\"u}sten}, \&
  {Wyrowski}}]{kama2016}
{Kama}, M., {Bruderer}, S., {van Dishoeck}, E.~F., {et~al.} 2016, \aap, 592,
  A83

\bibitem[{{Kama} {et~al.}(2020){Kama}, {Trapman}, {Fedele}, {Bruderer},
  {Hogerheijde}, {Miotello}, {van Dishoeck}, {Clarke}, \& {Bergin}}]{kama2020}
{Kama}, M., {Trapman}, L., {Fedele}, D., {et~al.} 2020, \aap, 634, A88

\bibitem[{{Krijt} {et~al.}(2016){Krijt}, {Ormel}, {Dominik}, \&
  {Tielens}}]{krijt2016a}
{Krijt}, S., {Ormel}, C.~W., {Dominik}, C., \& {Tielens}, A.~G.~G.~M. 2016,
  \aap, 586, A20

\bibitem[{{Krijt} {et~al.}(2018){Krijt}, {Schwarz}, {Bergin}, \&
  {Ciesla}}]{krijt2018}
{Krijt}, S., {Schwarz}, K.~R., {Bergin}, E.~A., \& {Ciesla}, F.~J. 2018, \apj,
  864, 78

\bibitem[{{Lambrechts} \& {Johansen}(2014)}]{lambrechts2014}
{Lambrechts}, M. \& {Johansen}, A. 2014, \aap, 572, A107

\bibitem[{{Lambrechts} {et~al.}(2019){Lambrechts}, {Morbidelli}, {Jacobson},
  {Johansen}, {Bitsch}, {Izidoro}, \& {Raymond}}]{lambrechts2019}
{Lambrechts}, M., {Morbidelli}, A., {Jacobson}, S.~A., {et~al.} 2019, \aap,
  627, A83

\bibitem[{{Le Gal} {et~al.}(2019){Le Gal}, {Brady}, {{\"O}berg}, {Roueff}, \&
  {Le Petit}}]{legal2019}
{Le Gal}, R., {Brady}, M.~T., {{\"O}berg}, K.~I., {Roueff}, E., \& {Le Petit},
  F. 2019, \apj, 886, 86

\bibitem[{{Lenz} {et~al.}(2019){Lenz}, {Klahr}, \& {Birnstiel}}]{lenz2019}
{Lenz}, C.~T., {Klahr}, H., \& {Birnstiel}, T. 2019, \apj, 874, 36

\bibitem[{{Long} {et~al.}(2017){Long}, {Herczeg}, {Pascucci}, {Drabek-Maunder},
  {Mohanty}, {Testi}, {Apai}, {Hendler}, {Henning}, {Manara}, \&
  {Mulders}}]{long2017}
{Long}, F., {Herczeg}, G.~J., {Pascucci}, I., {et~al.} 2017, \apj, 844, 99

\bibitem[{{Long} {et~al.}(2018){Long}, {Pinilla}, {Herczeg}, {Harsono},
  {Dipierro}, {Pascucci}, {Hendler}, {Tazzari}, {Ragusa}, {Salyk}, {Edwards},
  {Lodato}, {van de Plas}, {Johnstone}, {Liu}, {Boehler}, {Cabrit}, {Manara},
  {Menard}, {Mulders}, {Nisini}, {Fischer}, {Rigliaco}, {Banzatti}, {Avenhaus},
  \& {Gully-Santiago}}]{long2018}
{Long}, F., {Pinilla}, P., {Herczeg}, G.~J., {et~al.} 2018, \apj, 869, 17

\bibitem[{{Manara} {et~al.}(2016){Manara}, {Rosotti}, {Testi}, {Natta},
  {Alcal{\'a}}, {Williams}, {Ansdell}, {Miotello}, {van der Marel}, {Tazzari},
  {Carpenter}, {Guidi}, {Mathews}, {Oliveira}, {Prusti}, \& {van
  Dishoeck}}]{manara2016}
{Manara}, C.~F., {Rosotti}, G., {Testi}, L., {et~al.} 2016, \aap, 591, L3

\bibitem[{{Matr{\`a}} {et~al.}(2017){Matr{\`a}}, {MacGregor}, {Kalas}, {Wyatt},
  {Kennedy}, {Wilner}, {Duchene}, {Hughes}, {Pan}, {Shannon}, {Clampin},
  {Fitzgerald}, {Graham}, {Holland }, {Pani{\'c}}, \& {Su}}]{matra2017}
{Matr{\`a}}, L., {MacGregor}, M.~A., {Kalas}, P., {et~al.} 2017, \apj, 842, 9

\bibitem[{{McClure} {et~al.}(2016){McClure}, {Bergin}, {Cleeves}, {van
  Dishoeck}, {Blake}, {Evans}, {Green}, {Henning}, {{\"O}berg}, {Pontoppidan},
  \& {Salyk}}]{mcclure2016}
{McClure}, M.~K., {Bergin}, E.~A., {Cleeves}, L.~I., {et~al.} 2016, \apj, 831,
  167

\bibitem[{{McClure} \& {Dominik}(2019)}]{mcclure2019}
{McClure}, M.~K. \& {Dominik}, C. 2019, arXiv e-prints, arXiv:1910.07345

\bibitem[{{McElroy} {et~al.}(2013){McElroy}, {Walsh}, {Markwick}, {Cordiner},
  {Smith}, \& {Millar}}]{McElroy2013}
{McElroy}, D., {Walsh}, C., {Markwick}, A.~J., {et~al.} 2013, \aap, 550, A36

\bibitem[{{McKinnon} {et~al.}(2020){McKinnon}, {Richardson}, {Marohnic},
  {Keane}, {Grundy}, {Hamilton}, {Nesvorn{\'y}}, {Umurhan}, {Lauer}, {Singer},
  {Stern}, {Weaver}, {Spencer}, {Buie}, {Moore}, {Kavelaars}, {Lisse}, {Mao},
  {Parker}, {Porter}, {Showalter}, {Olkin}, {Cruikshank}, {Elliott},
  {Gladstone}, {Parker}, {Verbiscer}, {Young}, \& {New Horizons Science
  Team}}]{mckinnon2020}
{McKinnon}, W.~B., {Richardson}, D.~C., {Marohnic}, J.~C., {et~al.} 2020,
  Science, 367, aay6620

\bibitem[{{Miotello} {et~al.}(2019){Miotello}, {Facchini}, {van Dishoeck},
  {Cazzoletti}, {Testi}, {Williams}, {Ansdell}, {van Terwisga}, \& {van der
  Marel}}]{miotello2019}
{Miotello}, A., {Facchini}, S., {van Dishoeck}, E.~F., {et~al.} 2019, \aap,
  631, A69

\bibitem[{{Miotello} {et~al.}(2016){Miotello}, {van Dishoeck}, {Kama}, \&
  {Bruderer}}]{miotello2016}
{Miotello}, A., {van Dishoeck}, E.~F., {Kama}, M., \& {Bruderer}, S. 2016,
  \aap, 594, A85

\bibitem[{{Miotello} {et~al.}(2017){Miotello}, {van Dishoeck}, {Williams},
  {Ansdell}, {Guidi}, {Hogerheijde}, {Manara}, {Tazzari}, {Testi}, {van der
  Marel}, \& {van Terwisga}}]{miotello2017}
{Miotello}, A., {van Dishoeck}, E.~F., {Williams}, J.~P., {et~al.} 2017, \aap,
  599, A113

\bibitem[{{Misener} {et~al.}(2019){Misener}, {Krijt}, \&
  {Ciesla}}]{misener2019}
{Misener}, W., {Krijt}, S., \& {Ciesla}, F.~J. 2019, \apj, 885, 118

\bibitem[{{Morbidelli} {et~al.}(2014){Morbidelli}, {Szul{\'a}gyi}, {Crida},
  {Lega}, {Bitsch}, {Tanigawa}, \& {Kanagawa}}]{morbidelli2014}
{Morbidelli}, A., {Szul{\'a}gyi}, J., {Crida}, A., {et~al.} 2014, \icarus, 232,
  266

\bibitem[{{Nesvorn{\'y}}(2015)}]{nesvorny2015}
{Nesvorn{\'y}}, D. 2015, \aj, 150, 68

\bibitem[{{Nesvorn{\'y}} {et~al.}(2019){Nesvorn{\'y}}, {Li}, {Youdin}, {Simon},
  \& {Grundy}}]{nesvorny2019}
{Nesvorn{\'y}}, D., {Li}, R., {Youdin}, A.~N., {Simon}, J.~B., \& {Grundy},
  W.~M. 2019, Nature Astronomy, 3, 808

\bibitem[{{{\"O}berg} \& {Bergin}(2016)}]{oberg2016}
{{\"O}berg}, K.~I. \& {Bergin}, E.~A. 2016, \apjl, 831, L19

\bibitem[{{{\"O}berg} {et~al.}(2011){{\"O}berg}, {Murray-Clay}, \&
  {Bergin}}]{oberg2011}
{{\"O}berg}, K.~I., {Murray-Clay}, R., \& {Bergin}, E.~A. 2011, \apjl, 743, L16

\bibitem[{{Ormel}(2017)}]{ormel2017}
{Ormel}, C.~W. 2017, Astrophysics and Space Science Library, Vol. 445, {The
  Emerging Paradigm of Pebble Accretion}, ed. M.~{Pessah} \& O.~{Gressel}, 197

\bibitem[{{Padovani} {et~al.}(2018){Padovani}, {Ivlev}, {Galli}, \&
  {Caselli}}]{padovani2018}
{Padovani}, M., {Ivlev}, A.~V., {Galli}, D., \& {Caselli}, P. 2018, \aap, 614,
  A111

\bibitem[{{Pascucci} {et~al.}(2016){Pascucci}, {Testi}, {Herczeg}, {Long},
  {Manara}, {Hendler}, {Mulders}, {Krijt}, {Ciesla}, {Henning}, {Mohanty},
  {Drabek-Maunder}, {Apai}, {Sz{\H{u}}cs}, {Sacco}, \&
  {Olofsson}}]{pascucci2016}
{Pascucci}, I., {Testi}, L., {Herczeg}, G.~J., {et~al.} 2016, \apj, 831, 125

\bibitem[{{Pinilla} {et~al.}(2012){Pinilla}, {Benisty}, \&
  {Birnstiel}}]{pinilla2012}
{Pinilla}, P., {Benisty}, M., \& {Birnstiel}, T. 2012, \aap, 545, A81

\bibitem[{{Pinilla} {et~al.}(2020){Pinilla}, {Pascucci}, \&
  {Marino}}]{pinilla2020}
{Pinilla}, P., {Pascucci}, I., \& {Marino}, S. 2020, \aap, 635, A105

\bibitem[{{Pinte} {et~al.}(2019){Pinte}, {van der Plas}, {M{\'e}nard}, {Price},
  {Christiaens}, {Hill}, {Mentiplay}, {Ginski}, {Choquet}, {Boehler},
  {Duch{\^e}ne}, {Perez}, \& {Casassus}}]{pinte2019}
{Pinte}, C., {van der Plas}, G., {M{\'e}nard}, F., {et~al.} 2019, Nature
  Astronomy, 3, 1109

\bibitem[{{Piso} {et~al.}(2015){Piso}, {{\"O}berg}, {Birnstiel}, \&
  {Murray-Clay}}]{piso2015}
{Piso}, A.-M.~A., {{\"O}berg}, K.~I., {Birnstiel}, T., \& {Murray-Clay}, R.~A.
  2015, \apj, 815, 109

\bibitem[{{Powell} {et~al.}(2019){Powell}, {Murray-Clay}, {P{\'e}rez},
  {Schlichting}, \& {Rosenthal}}]{powell2019}
{Powell}, D., {Murray-Clay}, R., {P{\'e}rez}, L.~M., {Schlichting}, H.~E., \&
  {Rosenthal}, M. 2019, \apj, 878, 116

\bibitem[{{Prialnik} {et~al.}(2020){Prialnik}, {Barucci}, \&
  {Young}}]{prialnik2020}
{Prialnik}, D., {Barucci}, M.~A., \& {Young}, L. 2020, {The Trans-Neptunian
  Solar System}

\bibitem[{{Reboussin} {et~al.}(2015){Reboussin}, {Wakelam}, {Guilloteau},
  {Hersant}, \& {Dutrey}}]{reboussin2015}
{Reboussin}, L., {Wakelam}, V., {Guilloteau}, S., {Hersant}, F., \& {Dutrey},
  A. 2015, \aap, 579, A82

\bibitem[{{Rosotti} {et~al.}(2020){Rosotti}, {Teague}, {Dullemond}, {Booth}, \&
  {Clarke}}]{rosotti2020}
{Rosotti}, G.~P., {Teague}, R., {Dullemond}, C., {Booth}, R.~A., \& {Clarke},
  C. 2020, \mnras

\bibitem[{{Schwarz} {et~al.}(2016){Schwarz}, {Bergin}, {Cleeves}, {Blake},
  {Zhang}, {{\"O}berg}, {van Dishoeck}, \& {Qi}}]{schwarz2016}
{Schwarz}, K.~R., {Bergin}, E.~A., {Cleeves}, L.~I., {et~al.} 2016, \apj, 823,
  91

\bibitem[{{Schwarz} {et~al.}(2018){Schwarz}, {Bergin}, {Cleeves}, {Zhang},
  {{\"O}berg}, {Blake}, \& {Anderson}}]{schwarz2018}
{Schwarz}, K.~R., {Bergin}, E.~A., {Cleeves}, L.~I., {et~al.} 2018, \apj, 856,
  85

\bibitem[{{Schwarz} {et~al.}(2019){Schwarz}, {Bergin}, {Cleeves}, {Zhang},
  {{\"O}berg}, {Blake}, \& {Anderson}}]{schwarz2019}
{Schwarz}, K.~R., {Bergin}, E.~A., {Cleeves}, L.~I., {et~al.} 2019, \apj, 877,
  131

\bibitem[{{Semenov} \& {Wiebe}(2011)}]{semenov2011}
{Semenov}, D. \& {Wiebe}, D. 2011, \apjs, 196, 25

\bibitem[{{Semenov} {et~al.}(2006){Semenov}, {Wiebe}, \&
  {Henning}}]{semenov2006}
{Semenov}, D., {Wiebe}, D., \& {Henning}, T. 2006, \apjl, 647, L57

\bibitem[{{Shakura} \& {Sunyaev}(1973)}]{shakura1973}
{Shakura}, N.~I. \& {Sunyaev}, R.~A. 1973, \aap, 24, 337

\bibitem[{{Stammler} {et~al.}(2017){Stammler}, {Birnstiel}, {Pani{\'c}},
  {Dullemond}, \& {Dominik}}]{stammler2017}
{Stammler}, S.~M., {Birnstiel}, T., {Pani{\'c}}, O., {Dullemond}, C.~P., \&
  {Dominik}, C. 2017, \aap, 600, A140

\bibitem[{{Stern} {et~al.}(2019){Stern}, {Weaver}, {Spencer}, {Olkin},
  {Gladstone}, {Grundy}, {Moore}, {Cruikshank}, {Elliott}, {McKinnon},
  {Parker}, {Verbiscer}, {Young}, {Aguilar}, {Albers}, {Andert}, {Andrews},
  {Bagenal}, {Banks}, {Bauer}, {Bauman}, {Bechtold}, {Beddingfield}, {Behrooz},
  {Beisser}, {Benecchi}, {Bernardoni}, {Beyer}, {Bhaskaran}, {Bierson},
  {Binzel}, {Birath}, {Bird}, {Boone}, {Bowman}, {Bray}, {Britt}, {Brown},
  {Buckley}, {Buie}, {Buratti}, {Burke}, {Bushman}, {Carcich}, {Chaikin},
  {Chavez}, {Cheng}, {Colwell}, {Conard}, {Conner}, {Conrad}, {Cook}, {Cooper},
  {Custodio}, {Dalle Ore}, {Deboy}, {Dharmavaram}, {Dhingra}, {Dunn}, {Earle},
  {Egan}, {Eisig}, {El-Maarry}, {Engelbrecht}, {Enke}, {Ercol}, {Fattig},
  {Ferrell}, {Finley}, {Firer}, {Fischetti}, {Folkner}, {Fosbury}, {Fountain},
  {Freeze}, {Gabasova}, {Glaze}, {Green}, {Griffith}, {Guo}, {Hahn}, {Hals},
  {Hamilton}, {Hamilton}, {Hanley}, {Harch}, {Harmon}, {Hart}, {Hayes},
  {Hersman}, {Hill}, {Hill}, {Hofgartner}, {Holdridge}, {Hor{\'a}nyi},
  {Hosadurga}, {Howard}, {Howett}, {Jaskulek}, {Jennings}, {Jensen}, {Jones},
  {Kang}, {Katz}, {Kaufmann}, {Kavelaars}, {Keane}, {Keleher}, {Kinczyk},
  {Kochte}, {Kollmann}, {Krimigis}, {Kruizinga}, {Kusnierkiewicz}, {Lahr},
  {Lauer}, {Lawrence}, {Lee}, {Lessac-Chenen}, {Linscott}, {Lisse}, {Lunsford},
  {Mages}, {Mallder}, {Martin}, {May}, {McComas}, {McNutt}, {Mehoke}, {Mehoke},
  {Nelson}, {Nguyen}, {N{\'u}{\~n}ez}, {Ocampo}, {Owen}, {Oxton}, {Parker},
  {P{\"a}tzold}, {Pelgrift}, {Pelletier}, {Pineau}, {Piquette}, {Porter},
  {Protopapa}, {Quirico}, {Redfern}, {Regiec}, {Reitsema}, {Reuter},
  {Richardson}, {Riedel}, {Ritterbush}, {Robbins}, {Rodgers}, {Rogers}, {Rose},
  {Rosendall}, {Runyon}, {Ryschkewitsch}, {Saina}, {Salinas}, {Schenk},
  {Scherrer}, {Schlei}, {Schmitt}, {Schultz}, {Schurr}, {Scipioni}, {Sepan},
  {Shelton}, {Showalter}, {Simon}, {Singer}, {Stahlheber}, {Stanbridge},
  {Stansberry}, {Steffl}, {Strobel}, {Stothoff}, {Stryk}, {Stuart}, {Summers},
  {Tapley}, {Taylor}, {Taylor}, {Tedford}, {Throop}, {Turner}, {Umurhan}, {Van
  Eck}, {Velez}, {Versteeg}, {Vincent}, {Webbert}, {Weidner}, {Weigle},
  {Wendel}, {White}, {Whittenburg}, {Williams}, {Williams}, {Williams},
  {Winters}, {Zangari}, \& {Zurbuchen}}]{stern2019}
{Stern}, S.~A., {Weaver}, H.~A., {Spencer}, J.~R., {et~al.} 2019, Science, 364,
  aaw9771

\bibitem[{{Szul{\'a}gyi} {et~al.}(2014){Szul{\'a}gyi}, {Morbidelli}, {Crida},
  \& {Masset}}]{szulagyi2014}
{Szul{\'a}gyi}, J., {Morbidelli}, A., {Crida}, A., \& {Masset}, F. 2014, \apj,
  782, 65

\bibitem[{{Teague} {et~al.}(2019){Teague}, {Bae}, \& {Bergin}}]{teague2019}
{Teague}, R., {Bae}, J., \& {Bergin}, E.~A. 2019, \nat, 574, 378

\bibitem[{{Teague} {et~al.}(2018){Teague}, {Bae}, {Bergin}, {Birnstiel}, \&
  {Foreman-Mackey}}]{teague2018}
{Teague}, R., {Bae}, J., {Bergin}, E.~A., {Birnstiel}, T., \& {Foreman-Mackey},
  D. 2018, \apjl, 860, L12

\bibitem[{{Teague} {et~al.}(2016){Teague}, {Guilloteau}, {Semenov}, {Henning},
  {Dutrey}, {Pi{\'e}tu}, {Birnstiel}, {Chapillon}, {Hollenbach}, \&
  {Gorti}}]{teague2016}
{Teague}, R., {Guilloteau}, S., {Semenov}, D., {et~al.} 2016, \aap, 592, A49

\bibitem[{{Tielens} \& {Hagen}(1982)}]{tielens1982}
{Tielens}, A.~G.~G.~M. \& {Hagen}, W. 1982, \aap, 114, 245

\bibitem[{{Trapman} {et~al.}(2019){Trapman}, {Facchini}, {Hogerheijde}, {van
  Dishoeck}, \& {Bruderer}}]{trapman2019}
{Trapman}, L., {Facchini}, S., {Hogerheijde}, M.~R., {van Dishoeck}, E.~F., \&
  {Bruderer}, S. 2019, \aap, 629, A79

\bibitem[{{Umebayashi} \& {Nakano}(1981)}]{umebayashi1981}
{Umebayashi}, T. \& {Nakano}, T. 1981, \pasj, 33, 617

\bibitem[{{Walsh} {et~al.}(2010){Walsh}, {Millar}, \& {Nomura}}]{walsh2010}
{Walsh}, C., {Millar}, T.~J., \& {Nomura}, H. 2010, \apj, 722, 1607

\bibitem[{{Willacy} {et~al.}(2006){Willacy}, {Langer}, {Allen}, \&
  {Bryden}}]{willacy2006}
{Willacy}, K., {Langer}, W., {Allen}, M., \& {Bryden}, G. 2006, \apj, 644, 1202

\bibitem[{{Williams} \& {Best}(2014)}]{williams2014}
{Williams}, J.~P. \& {Best}, W. M.~J. 2014, \apj, 788, 59

\bibitem[{{Wyatt}(2020)}]{wyatt2020}
{Wyatt}, M. 2020, {Extrasolar Kuiper belts}, ed. D.~{Prialnik}, M.~A.
  {Barucci}, \& L.~{Young}, 351--376

\bibitem[{{Xu} {et~al.}(2017){Xu}, {Bai}, \& {{\"O}berg}}]{xu2017}
{Xu}, R., {Bai}, X.-N., \& {{\"O}berg}, K. 2017, \apj, 835, 162

\bibitem[{{Zhang} {et~al.}(2017){Zhang}, {Bergin}, {Blake}, {Cleeves}, \&
  {Schwarz}}]{zhang2017}
{Zhang}, K., {Bergin}, E.~A., {Blake}, G.~A., {Cleeves}, L.~I., \& {Schwarz},
  K.~R. 2017, Nature Astronomy, 1, 0130

\bibitem[{{Zhang} {et~al.}(2019){Zhang}, {Bergin}, {Schwarz}, {Krijt}, \&
  {Ciesla}}]{zhang2019}
{Zhang}, K., {Bergin}, E.~A., {Schwarz}, K., {Krijt}, S., \& {Ciesla}, F. 2019,
  \apj, 883, 98

\bibitem[{{Zhang} {et~al.}(2020{\natexlab{a}}){Zhang}, {Bosman}, \&
  {Bergin}}]{zhang2020b}
{Zhang}, K., {Bosman}, A.~D., \& {Bergin}, E.~A. 2020{\natexlab{a}}, \apjl,
  891, L16

\bibitem[{{Zhang} {et~al.}(2020{\natexlab{b}}){Zhang}, {Schwarz}, \&
  {Bergin}}]{zhang2020a}
{Zhang}, K., {Schwarz}, K.~R., \& {Bergin}, E.~A. 2020{\natexlab{b}}, \apjl,
  891, L17

\end{thebibliography}

\end{document}